\documentstyle[12pt]{article}
\makeatletter
\def\bigans{y }
\read-1 to\answ
\ifx\answ\bigans
\ifcase\@ptsize
 \font\tenmsa=msam10
 \font\sevenmsa=msam7
 \font\fivemsa=msam5
 \font\tenmsb=msbm10
 \font\sevenmsb=msbm7
 \font\fivemsb=msbm5
\or
 \font\tenmsa=msam10 scaled \magstephalf
 \font\sevenmsa=msam8
 \font\fivemsa=msam6
 \font\tenmsb=msbm10 scaled \magstephalf
 \font\sevenmsb=msbm8
 \font\fivemsb=msbm6
\or
 \font\tenmsa=msam10 scaled \magstep1
 \font\sevenmsa=msam8
 \font\fivemsa=msam6
 \font\tenmsb=msbm10 scaled \magstep1
 \font\sevenmsb=msbm8
 \font\fivemsb=msbm6
\fi

\newfam\msafam
\newfam\msbfam
\textfont\msafam=\tenmsa  \scriptfont\msafam=\sevenmsa
  \scriptscriptfont\msafam=\fivemsa
\textfont\msbfam=\tenmsb  \scriptfont\msbfam=\sevenmsb
  \scriptscriptfont\msbfam=\fivemsb

\def\hexnumber@#1{\ifnum#1<10 \number#1\else
 \ifnum#1=10 A\else\ifnum#1=11 B\else\ifnum#1=12 C\else
 \ifnum#1=13 D\else\ifnum#1=14 E\else\ifnum#1=15 F\fi\fi\fi\fi\fi\fi\fi}

\def\msa@{\hexnumber@\msafam}
\def\msb@{\hexnumber@\msbfam}

\catcode`\@=12
\def\ZZ{{\Bbb Z}}
\def\CC{{\Bbb C}}

\else
\def\CC{I\!\!\!\!C}
\def\ZZ{Z\!\!\!Z}
		\fi

\catcode`@=11
\def\citen#1{\if@filesw \immediate\write \@auxout {\string\citation{#1}}\fi%
\@tempcntb\m@ne \let\@h@ld\relax \def\@citea{}%
\@for \@citeb:=#1\do {\@ifundefined {b@\@citeb}%
    {\@h@ld\@citea\@tempcntb\m@ne{\bf ?}%
    \@warning {Citation `\@citeb ' on page \thepage \space its cite}}%
    {\@tempcnta\@tempcntb \advance\@tempcnta\@ne
    \setbox\z@\hbox\bgroup\ifcat0\csname b@\@citeb \endcsname \relax
    \egroup \@tempcntb\number\csname b@\@citeb \endcsname \relax
    \else \egroup \@tempcntb\m@ne \fi \ifnum\@tempcnta=\@tempcntb
    \ifx\@h@ld\relax \edef \@h@ld{\@citea\csname b@\@citeb\endcsname}%
    \else \edef\@h@ld{\hbox{--}\penalty\@highpenalty
    \csname b@\@citeb\endcsname}\fi
    \else \@h@ld\@citea\csname b@\@citeb \endcsname \let\@h@ld\relax \fi}%
\def\@citea{,\penalty\@highpenalty\hskip.13em plus.13em minus.13em}}\@h@ld}
\def\@citex[#1]#2{\@cite{\citen{#2}}{#1}}%
\def\@cite#1#2{\leavevmode\unskip\ifnum\lastpenalty=\z@\penalty\@highpenalty\fi%
   $^{\scriptscriptstyle \multiply\@highpenalty 3 \mbox{\rm\scriptsize#1%
  \if@tempswa,\penalty\@highpenalty\ #2\fi}}$}   %
\def\dciten#1{\if@filesw \immediate\write \@auxout {\string\citation{#1}}\fi%
\@tempcntb\m@ne \let\@h@ld\relax \def\@dcitea{}%
\@for \@dciteb:=#1\do {\@ifundefined {b@\@dciteb}%
    {\@h@ld\@dcitea\@tempcntb\m@ne{\bf ?}%
    \@warning {line Citation `\@dciteb ' on page \thepage \space its dcite}}%
    {\@tempcnta\@tempcntb \advance\@tempcnta\@ne
    \setbox\z@\hbox\bgroup\ifcat0\csname b@\@dciteb \endcsname \relax
    \egroup \@tempcntb\number\csname b@\@dciteb \endcsname \relax
    \else \egroup \@tempcntb\m@ne \fi \ifnum\@tempcnta=\@tempcntb
    \ifx\@h@ld\relax \edef \@h@ld{\@dcitea\csname b@\@dciteb\endcsname}%
    \else \edef\@h@ld{\hbox{--}\penalty\@highpenalty
    \csname b@\@dciteb\endcsname}\fi
    \else \@h@ld\@dcitea\csname b@\@dciteb \endcsname \let\@h@ld\relax \fi}%
\def\@dcitea{,\penalty\@highpenalty\hskip.13em plus.13em minus.13em}}\@h@ld}
\def\@dcitex[#1]#2{\@dcite{\dciten{#2}}{#1}}%
\def\@dcite#1#2{\leavevmode\ifnum\lastpenalty=\z@\penalty%
  \@highpenalty\fi%
 {\multiply\@highpenalty 3
   {\rm [#1]\if@tempswa,\penalty\@highpenalty  #2\fi}}}   %
\def\dcite{\@ifnextchar [{\@tempswatrue\@dcitex}{\@tempswafalse\@dcitex[]}}
\catcode`@=12

\renewcommand{\theequation}{\thesection.\arabic{equation}}

\def\onward{\addtocounter{section}{1} \setcounter{equation}{0}
                                           \setcounter{subsection}{0} }
\def\subonward{\addtocounter{subsection}{1} }

\outer\def\topic#1{\par{\vskip0pt plus.2\vsize\penalty-250
                        \vskip0pt plus-.2\vsize\bigskip\vskip\parskip
      \onward\message{#1}\vspace{0.5cm}
      \leftline{{\large\bf \thesection. #1}}\nobreak\smallskip  \noindent}}
\outer\def\subtopic#1{\par{\vskip0pt plus.05\vsize\penalty-250
                           \vskip0pt plus-.05\vsize\bigskip\vskip\parskip
   \subonward \vspace{0.25cm}
           \leftline{{\it \thesubsection\/ #1}}\nobreak\noindent}}

\def\smultab#1#2#3#4#5#6#7{\put (0,4){\line(1,0){#1}}
                    \multiput(0,3)(1,0){#1}{\line(1,0){1}}
                    \multiput(1,3)(1,0){#1}{\line(0,1){1}}
                    \multiput(0,2)(1,0){#2}{\line(1,0){1}}
                    \multiput(1,2)(1,0){#2}{\line(0,1){1}}
                    \multiput(0,1)(1,0){#3}{\line(1,0){1}}
                    \multiput(1,1)(1,0){#3}{\line(0,1){1}}
                    \multiput(0,0)(1,0){#4}{\line(1,0){1}}
                    \multiput(1,0)(1,0){#4}{\line(0,1){1}}
                    \multiput(0,-1)(1,0){#5}{\line(1,0){1}}
                    \multiput(1,-1)(1,0){#5}{\line(0,1){1}}
                    \multiput(0,-2)(1,0){#6}{\line(1,0){1}}
                    \multiput(1,-2)(1,0){#6}{\line(0,1){1}}
                          \put (0,4){\line(0,-1){#7}}}

\def\young#1#2#3#4#5#6#7{\begin{picture}(#1,#7)(0,3)
                \thicklines \smultab#1#2#3#4#5#6#7 \end{picture}}

 \def\fund{\young1000001}
\def\yfund#1{\raise #1mm \hbox{\young1100002}}

\newcommand{\eq}{\begin{equation}}
\newcommand{\en}{\end{equation}}
\def\ie{{\it i.e.}}
\def\Mabhat{{M_{ab}}^{\bbox}}
\def\Labhat{{L_{ab}}^{\bbox}}

\def\col{m}

\def\row{\ell}

\def\tchi{K}
\def\hchi{{\raise 0.8mm \hbox{$\chi$}}}
\def\bbox{\hat{c}}
\def\colK{\cK}

\def\cI{{\cal I}}

\def\cR{{\cal R}}

\def\cF{{\cal F}}
\def\cfT{{\cal T}}
\def\cS{{\cal S}}
\def\cN{{\cal N}}

\def\cG{{\cal G}}
\def\cK{{\cal K}}

\def\cC{{\cal C}}
\def\cT{{\cal T}}

\def\half{{\scriptstyle \frac{1}{2}}}
\def\sst{\scriptscriptstyle}
\def\sct{\scriptstyle}
\def\ds{\displaystyle}
\def\tx{\textstyle}

\def\Nabc{{N_{ab}}^{c}}

\def\ta{\widetilde{a}}

\def\vev#1{\langle #1 \rangle}

\def\char{{\rm char}}

\def\wt#1{\widetilde{#1}}

\def\sun{{\hbox{\sc su}}(N)}
\def\smcp{{\rm\footnotesize p}}
\def\Sp{\hbox{\sc s\smcp}}

\def\son{{\hbox{\sc so}}(N)}
\def\sunless{{\hbox{\sc su}}(n)}
\def\sunmorek{{\hbox{\sc su}}(n+1)_k}
\def\sunk{{\hbox{\sc su}}(N)_{K}}
\def\su{\hbox{\sc su}}
\def\sukn{{\hbox{\sc su}}(K)_{N}}
\def\sonk{{\hbox{\sc so}}(N)_{K}}

\def\spnk{{\hbox{\sc s\smcp}}(N)_{K}}
\def\spsnk{{\hbox{\sc s\smcp}}(n)_{k}}
\def\spkn{{\hbox{\sc s\smcp}}(K)_{N}}
\def\qbox#1{\quad\hbox{#1}\quad}
\def\smbullet{{\footnotesize $\bullet$}}
\def\bfid{\rlap{\bf 1}\phantom{\setlength{\unitlength}{0.2cm} \fund}}
\def\ijkl#1{\rlap{$#1$}\phantom{\setlength{\unitlength}{0.2cm} \fund}}
\def\lph{\hbox{\phantom{$-($}}}
\def\rph{\hbox{\phantom{$)$}}}
\def\lphm{\hbox{\phantom{$-$}}}
\def\LG{Landau-Ginzburg }
\def\bsub#1{\hbox{\footnotesize $#1$}}
\def\sodnk{\hbox{\sc so}(2n+1)_{2k+1} }
\def\sodnK{\hbox{\sc so}(2n+1)_{K} }
\def\sodkn{\hbox{\sc so}(2k+1)_{2n+1}}

\def\sodn{\hbox{\sc so}(2n+1) }
\def\cV{{\cal V}}

\def\hep#1#2#3#4{hepth$\bullet$}

\begin{document}
\setlength{\unitlength}{0.25cm}
\thispagestyle{empty}

\hfill           \begin{tabular}{l} {\bf \hep-th/9310082} \\
                                    {\sf BRX-TH--348} \\
                                   \end{tabular}
\vspace{1.7cm}

\setcounter{footnote}{1}
\begin{center}
{\LARGE  Integrable $\cN = 2$ Landau-Ginzburg Theories \\[0.2cm]
           from Quotients of Fusion Rings\footnote{Supported in part by
               the DOE under grant DE-FG02-92ER40706}}\\[1.5cm]

{\large \sl Eli J. Mlawer, Harold A. Riggs, and Howard J. Schnitzer }

\vspace{0.5cm}
{ \sl
\begin{tabular}{c}
   Department of Physics \\
   Brandeis University \\
   Waltham, MA 02254
\end{tabular}  }

\end{center}

\vfill
\begin{center}
{\sc Abstract}
\end{center}

\begin{quotation}

The discovery of integrable $\cN=2$ supersymmetric \LG
theories whose chiral rings are fusion rings suggests a close connection
between fusion rings, the related \LG superpotentials,
and $\cN=2$ quantum integrability. We examine this connection by finding the
natural $\sonk$ analogue of the construction that produced
the superpotentials with $\spnk$ and $\sunk$ fusion rings as chiral rings.
The chiral rings of the new superpotentials are {\em not} directly the
fusion rings of {\em any} conformal field theory, although they are natural
quotients of the tensor subring of the $\sonk$ fusion ring.

The new superpotentials yield solvable (twisted $\cN=2$) topological field
theories. We obtain the integer-valued correlation functions
as sums of $\sonk$ Verlinde dimensions by expressing
the correlators as fusion
residues. The $\sodnk$ and $\sodkn$ related
topological \LG theories are isomorphic,
despite being defined via quite different superpotentials.

\end{quotation}
\vfill
{\sf September 1993}  \hfill

\setcounter{page}{0}
\newpage
\setcounter{page}{1}

\topic{Introduction}

The quantum field theories on worldsurfaces that possess $\cN=2$
supersymmetry continue to yield new surprises and find new areas of
application,\cite{harvone,harvtwo,poly,morefw,spect} even
after several years of intense scrutiny.\cite{revwarner}
In addition to playing crucial roles in attempts to
extract testable physics from string theory,\cite{phenom} in the discovery
of mirror symmetry in algebraic geometry,\cite{mirror}
and in the understanding of how matter couples to quantum gravity
in two dimensions,\cite{witten,dykg,kkli,vvd,grav} such theories have
recently found realizations
in the critical behaviour of exactly solvable lattice
models\cite{latwarner} and have even led to the solution of certain
long-standing problems in (experimentally realizable) two-dimensional polymer
physics.\cite{poly}

Those $\cN=2$ field theories that are characterized by a
superpotential, the $\cN=2$ \LG theories,\cite{martharv,lerchewarn}
are important not least because many of their properties are explicitly
calculable. Special one-parameter families of
superpotentials have been found that smoothly  deform $\cN=2$
superconformal theories into massive $\cN=2$ supersymmetric---and
quantum integrable\cite{intzero,intone,intwo}---field theories.
In many such cases the chiral ring
(the ring of the chiral, $\cN=2$ primary fields) of the superconformal
theory\cite{lerchewarn} deforms to a ring isomorphic to the fusion ring of a
rational conformal field theory. For example, the superpotentials whose
chiral rings are Grassmannian cohomology rings\cite{lerchewarn} can be
perturbed in different directions to obtain massive (apparently integrable)
supersymmetric theories with chiral rings that are isomorphic to
the $\sunk$ and $\spnk$ current algebra fusion rings.\cite{gepner,brsym}
These particular fusion rings have properties
(such as automorphisms generated
by simple currents) that an arbitrary fusion ring\cite{ffrev} does
not possess. This leaves open the possibility that
polynomial potentials which yield integrable theories only arise
from fusion rings with such extra structure.
Since {\em any} finite set of distinct points in an $n$-dimensional
complex vector space can be represented as
the critical points of some polynomial in $n$
variables,\cite{saclay,aharoni}
the existence itself of a polynomial whose critical points
reproduce the solutions of a fusion ring
\setcounter{footnote}{1}
cannot be a crucial factor for integrability.\footnote{It
remains unclear in which cases
potentials exist that flow smoothly (\ie, without changing
the number of superconformal chiral primary fields) to
a rational fusion ring.} The direct construction of covering-space
potentials\cite{cresc} faces the problem of translation to
physical variables.
The interesting question concerning such integrable deformations remains:
What is the crucial principle behind the magic of these special
polynomial potentials?  A promising direction is the
connection between the structure of certain {\it graph rings}\cite{saclay}
and integrability.

In this paper we construct a new family of  massive
superpotentials. They are natural $\sonk$ parallels of the
$\sunk$ and $\spnk$ related potentials and have, we expect, a good
chance of leading to integrable theories.
However, the associated chiral rings are not
the fusion rings of any conformal field theory.
While these chiral rings are related to the
$\sonk$ fusion rings, they reproduce only a quotient of
an $\sonk$ fusion subring.  Further,
unlike the $\sunk$ and $\spnk$ fusion rings, they
do not possess a simple-current automorphism.
This suggests that simple-current automorphisms are not
prerequisites for natural (possibly integrable) Landau-Ginzburg constructions.
We will only consider the case $\sodnk$ in this paper.
A similar but more involved construction is possible for $\sonk$ for any
$N$ and $K$ with similar results, the details of which
will be presented elsewhere.

In section two we use the idea of Young tableau transposition
symmetry to define quotient rings of the tensor subring of the
$\sodn$ representation ring and establish a connection with
the $\sodnk$ fusion ring.
In section three we show that although these rings are
fusion rings, they cannot, in general,
be the fusion rings of any conformal field theory.
In section four we construct potentials whose local rings
exactly reproduce these quotient rings. In section five,
we use these potentials as the superpotentials of massive $\cN=2$
supersymmetric field theories, and find that
the superpotential related to $\sodnk$ is a new deformation of the
superpotential whose chiral ring is the homology intersection
ring of the Grassmannian
$$
        { U(n+k) \over U(n) \otimes U(k)} \; .
$$
This same Grassmannian superpotential
can be perturbed in other directions to obtain
\LG models with chiral rings isomorphic to
the ${\hbox{\sc su}}(n+1)_{k}$ and ${\hbox{\sc s\smcp}}(n)_{k}$
current algebra fusion rings. As an immediate application, we consider the
twisted $\cN=2$ topological
\LG models\cite{twistop,witten,dykg,kkli,vvd} that can be
obtained from these $\sodnk$ related
superpotentials, and write the correlation
functions on any genus in terms of
the Verlinde numbers of the $\sodnk$ fusion
ring. We also show that the twisted versions of the $\sodnk$ and
$\sodkn$ based topological \LG theories are
identical (on surfaces of any genus).
In the concluding section we comment on the
pattern formed by the fusion-related potentials discovered to date.

\topic{Transposition Ideals and Cominimal Quotients}

First we will exhibit $n$ independent generating relations for a
sequence of ideals in the ring of polynomials in the $n$
fundamental tensor characters $\hchi_i$ of $\sodn$,
$\ZZ[\hchi_1,\ldots, \hchi_n]$. Then we write the
values of the $\hchi_i$ that satisfy the generating relations
of the $k^{\rm th}$ ideal as ratios of $\sodnk$ modular transformation
matrix elements. This is possible due to the close connection of the
related quotient ring,
$$
\cR_{n,k}  = {\ZZ[\hchi_1,\ldots, \hchi_n] \over \cI_k} \; ,
$$
to the $\sodnk$ fusion algebra.

\subtopic{The Transposition Ideals}

The irreducible representations of the complex Lie algebras $\sodn$
(\ie, $B_n$ for $n\geq 3$, $C_2$ for $n=2$, and $A_1$ for $n=1$)
are naturally classified by the Young tableaux with at most
$n$ rows. The row lengths of the tableau for a given
irreducible representation are related to the Dynkin indices
of its highest weight by
\eq
\begin{array}{cclr}
\ds \row_{j} & = & \ds \sum_{i=j}^{n-1} a_{i} + {\tx \frac{1}{2}} a_{n}
              & \quad 1 \leq j \leq n-1 \\[0.4cm]
    \ds  \row_{n} & = & \frac{1}{2} a_{n} \; . &
\end{array}
\label{dynkind}
\en
We will only consider tensor representations
(for $n=1$ and $n=2$ only real representations) in the following so that
the $\row_i$ will always be integral; the Dynkin index of the
last root, $a_n$, is then even.
For $n=2$, $a_2$ labels, contrary to custom,
\setcounter{footnote}{1}
the short root.\footnote{Note that the $\hbox{\sc SO}(5)$ tableau
(\ref{dynkind}) for
a given representation of $C_2$ differs significantly from the
standard $\hbox{\sc S\rm p}(2)$ tableau for the same representation.
Similarly, any $\hbox{\sc SO}(3)$ tableau has half the cells of the
$\hbox{\sc SU}(2)$ tableau for the same $A_1$ representation.}
The $\sodn$ tensor language allows an elegant and uniform description
of all topics dealt with in this paper for all $n$.

Let $\tchi_{i}$ denote the character of the representation associated
with the single\-row tableau of length $\row_1 = i$.
Then the character of an arbitrary representation,
specified by a tableau $a$ with $\col_1$ non-zero row lengths $\row_i$,
is given by  the $\col_1 \times \col_1$ determinant\cite{fh}
\eq
  \char_{\bsub{a}} = {\tx \frac{1}{2}}
 \det \left| \tchi_{\row_i -i+j}  + \tchi_{\row_i -i + 2 - j} \right|
\label{transdet}
\en
where $i,j = 1, \ldots, \col_1$, and $K_j = 0$ for
$j<0$. This determinant is well-defined
for {\em any} tableau, so
that we can always refer to the character of a given tableau, even if
that tableau does not correspond to a representation of $\sodn$.
In particular, let $\hchi_i$ denote the single column
tableau of height $\col_1 = i$ for $i = 0, \ldots, \infty$.
If $1 \leq i \leq n$ then $\hchi_i$
denotes the character of the $i^{{\rm th}}$ fundamental tensor
representation. It is important to
realize that the $\hchi_i$ with $i > n$ do not all vanish, but
instead satisfy\cite{cking}
\eq
  \begin{array}{rl}
      \hchi_{n+j} - \hchi_{n-j+1} = 0 & \rlap{\qquad $1\leq j \leq n+1$} \\
   \hchi_{n+j} = 0 &  \rlap{\qquad  $\hbox{\phantom{$1\leq$}}
              \;   j   > n+1 \; .$}
          \end{array}
\label{rmodgen}
\en
This follows from the explicit expression for the $K_i$ as
traces in the relevant representations as well as the definition of
the $\hchi_i$ in (\ref{transdet}).
These identities allow one to write the
character of any tableau with more than $n$ rows  in terms
of a character with at most $n$ rows, as follows.

With $\hchi_0 = 1$ (the character of the identity representation) and
$\hchi_j = 0$ for $j<0$, the character of any tableau $a$ with
$\row_1$ non-zero column lengths $\col_i$ is given by the
$\row_1 \times \row_1$ determinant\cite{fh}
\eq
   \char_{\bsub{a}} =  {\tx \frac{1}{2}}
    \det \left| \hchi_{\col_i -i+j}  + \hchi_{\col_i -i + 2 - j} \right|
\label{basdet}
\en
with  $i,j= 1, \ldots, \row_1$. Note that formula (\ref{basdet})
transforms into formula (\ref{transdet}) under the interchange of
rows and columns, as long as the identities (\ref{rmodgen}) are
disregarded. Upon using these latter identities in the determinant
(\ref{basdet}) one obtains
the character associated with any tensor representation (or with any
tableau) as a polynomial
in the fundamental characters
$\hchi_i$ with $i=0,1,\ldots, n$. It is
remarkable that, as a consequence of (\ref{rmodgen}) and
(\ref{basdet}), the character of a tableau with more
than $n$ rows can be transformed (up to sign) into
the character of a {\em single} tableau with at most $n$ rows, according to
a simple rule.\cite{cking} This {\it rank modification rule}, which implements
certain products of Weyl group reflections via
the removal of strips of cells on the tableau
boundary, embodies the implications of (\ref{rmodgen})
for the characters of arbitrary tableaux.

The product of characters defined by the product of the polynomials
(\ref{basdet}) without imposition of the identities (\ref{rmodgen})
is given by\cite{cking}
\eq
 \char_{\bsub{a}}\; \char_{\bsub{b}} = \sum_{c} {T_{ab}}^{c}\; \char_{\bsub{c}}
 \equiv \sum_{d} \char \{(a/d) \cdot (b/d) \}  \; .
\label{mixal}
\en
The raised dot indicates the product
of tableaux that is given by the Littlewood-Richardson
rule, and
$$
  (a/d) = \sum_{e} {L_{d e}}^{a} \; e
$$
denotes the formal sum of all tableaux $e$ (with Littlewood-Richardson
multiplicity ${L_{d e}}^{a}$) such that $a \in d \cdot e$.
Equation (\ref{mixal}) represents
the tableau product in the ring of polynomials in the infinite set of
variables $\hchi_i$, $i=0,1,\ldots \infty$.
With $\wt{a}$ denoting the transpose
of the tableau $a$ (\ie, $\row_i(\wt{a}) = \col_i(a)$), the transposition
symmetry of the Littlewood-Richardson multiplicities\cite{lrtran}
$$
        {L_{\, \wt{a\mathstrut}\, \wt{b\mathstrut} }}^{\;\, \wt{c}} =
              {L_{\vphantom{\wt{b\mathstrut}}a b}}^{c}
$$
and the aformentioned transposition symmetry of the determinant
formulae guarantee that the transposition symmetry continues to hold
for the tableau multiplicities in (\ref{mixal}), so that
\eq
      {T_{\; \wt{a\mathstrut}\, \wt{b\mathstrut} }}^{\;\, \wt{c}} =
              {T_{\;\vphantom{\wt{b\mathstrut}} a b}}^{c}   \; .
\label{transym}
\en
Since the characters of tableaux with more
than $n$ rows often appear in the product (\ref{mixal}),
the rank modification rules that implement the identities (\ref{rmodgen})
must be imposed in order to obtain the representation ring of the tensor
representations of $\sodn$ from (\ref{mixal}).  Once this is done, the free
ring of polynomials in the fundamental characters
$\hchi_{0}, \hchi_{1},\ldots,\hchi_{n}$ generates the
tensor ring of $\sodn$, with the translation between polynomials in the
$\hchi_{i}$ and tableaux given by the determinant formula in (\ref{basdet}).
However, the transposition symmetry
just mentioned is lost.

It is interesting that the generators of the local
rings of the $\sunk$ and $\spnk$ fusion potentials are transposes
(upon interchange of $N$ and $K$)
of the generators of the respective rank modification rules
for these Lie algebras.
Therefore, it is natural to consider the ideals $\cI_k$ of
the $\sodn$ tensor representation ring that are generated by the relations
\eq
     \tchi_{k+j} - \tchi_{k-j+1} = 0   \rlap{$\qquad 1\leq j \leq n\; .$}
\label{genvan}
\en
These generators are (upon interchange of $n$ and $k$)
exact tableau transposes of (a subset of) the identities (\ref{rmodgen})
satisfied by the $\hchi_i$. We will call these ideals
the {\it transposition ideals}. Since (\ref{genvan}) does {\em not}
hold for all $j>n$ (see equation~\ref{onerow} in the appendix and
the comment that follows there),
it is nontrivial that the quotient rings
\eq
    \cR_{n,k} = {\ZZ[\hchi_1,\ldots,\hchi_n] \over \cI_k}
\label{quotring}
\en
exhibit a restoration of the transposition symmetry (\ref{transym})
that is lost when the rank modification rules are imposed.

\subtopic{The Cominimal Quotients}

We will now describe certain quotients of the polynomial
ring $\ZZ[\hchi_1,\ldots,\hchi_n]$ that are directly related to
the $\sodnk$ fusion rings and their simple currents.

The standard basis elements $\phi_a$
for the fusion ring associated with conformal\-scalar
fields carrying representations of the level $K$ untwisted
affine Lie algebra $\sodnK$ (\ie, $A_1^{(1)}$ at level $2K$ for $n=1$,
$C_2^{(1)}$ at level $K$
for $n=2$, and $B_n^{(1)}$ at level
$K$ for $n\geq 3$) are labelled by
the irreducible and integrable highest weight representations of $\sodnK$.
These representations are classified by the tableaux with at most $n$ rows
whose (possibly half-integral)
row lengths also satisfy $\row_1 + \row_2 \leq K$.
We will specify the tensor subring of the
$\sodnk$ fusion ring by
$\cfT_{n, k}$ and will call $k$ the {\it reduced} level.

The non-negative integers $\Nabc$ that define the fusion ring product
\eq
           \phi_a \star \phi_b = \sum_c \Nabc\;  \phi_c \; ,
\label{fusrule}
\en
are related to the modular transformation matrix elements
$S_{ab}$, which are all real for $\sodnK$, by Verlinde's
sum\cite{verlind}
\eq
              {N_{ab}}^c = \sum_r
         {S_{ar} S_{br} S_{cr} \over S_{0r}}
\label{verlinde}
\en
over {\em all} level $K$ integrable representations.
This formula implies that the fusion coefficients are
completely determined by
an extension of the  Speiser algorithm for calculating
ordinary Kronecker products.\cite{exspeiser} The extended
algorithm exhibits a WZW fusion ring as a quotient of the
relevant representation ring by a set of relations between
characters related to certain products of affine Weyl group reflections.
We will refer to these character relations as the
{\it fusion modification rules}\cite{cummins}
and the ideal they generate, $\cF_k$, as the fusion ideal. This means that
\eq
         \cfT_{n,k} = {\ZZ[\hchi_1,\ldots,\hchi_n] \over \cF_k} \; .
\label{tenssub}
\en

The $\sodnk$ fusion ring has a nontrivial
automorphism associated with the $\ZZ_2$ automorphism of the
extended $\sodn$ Dynkin diagram.\cite{firf}
This diagram automorphism induces a
map $\sigma$ between the integrable
highest weight representations that label the basis
elements of the fusion ring.
In terms of the tableau row lengths that label these representations,
this map leaves unchanged all row lengths except the first, which
transforms according to
\eq
            \row_1 (\sigma(a)) = K - \row_1(a) \; .
\label{tabcomin}
\en
We will call the quotient of $\cfT_{n,k}$ by the
ideal $\cC_k$ composed of the entire set of identities
\eq
           \phi_{\sigma(t)} - \phi_t = 0
\label{comin}
\en
where $t$ is any tableaux with $\row_1+\row_2\leq K$,
the
\setcounter{footnote}{1}
{\it cominimal\/}\footnote{The term arises from the relation
between the WZW fields that generate simple-currents and cominimal
highest weights.\cite{firf}}
quotient of $\cfT_{n,k}$.
The multiplicities that define the tableau product
in this quotient ring are given by
\eq
      {M_{ab}}^{c} = {N_{ab}}^{c} + {N_{ab}}^{\sigma(c)} \; .
\label{quotmult}
\en
Since the representations of $\sodn$ are self-conjugate, and since
cominimally equivalent representations have just been equated in (\ref{comin}),
the cominimal quotient of $\cfT_{n,k}$ has no apparent
non-trivial automorphisms. However, if $n=k$, there is a hidden
automorphism given by tableau transposition, as will
become apparent.

We will label the basis elements of this cominimal quotient ring by
the $n$ row tableaux with $\row_1 \leq k$. Then we may
obtain the quotient-ring product by using the
identities (\ref{comin}) to eliminate any
tableau with first row length $\row_1$ greater than the reduced level $k$
from  fusion rule (\ref{fusrule}).
The two ideals commute so that we can write
\eq
{\cfT_{n,k} \over \cC_k} =
{\ZZ[\hchi_1,\ldots,\hchi_n] \over \cF_k \cdot \cC_k}  \; .
\label{comquot}
\en

If $a$ denotes any basis tableau of $\cfT_{n,k}/\cC_k$, then
(since $\row_1(\widetilde{a})\leq n$ and $\col_1(\widetilde{a})\leq k$)
$\widetilde{a}$ is always a basis tableau of $\cfT_{k,n}/\cC_n$.
With tableau transposition considered as a map between
$\cfT_{n,k}/\cC_k$ and $\cfT_{k,n}/\cC_n$ (as well as an operation on
tableaux), the fact that $\widetilde{\widetilde{a}}=a$, means that
transposition gives a one-to-one correspondence between elements of
$\cfT_{n,k}/\cC_k$ and $\cfT_{k,n}/\cC_n$. This correspondence is also an
isomorphism: It was shown some time ago\cite{dual4} that the
$\sodnk$ and $\sodkn$ fusion multiplicities are related by
\eq
   \left( {N_{\vphantom{\wt{b\mathstrut}}a b}}^{c}
            \right)_{\sodnk} =
 \left( {N_{\, \wt{a\mathstrut}\,%
   \wt{b\mathstrut} }}^{\;\, \sigma^{\Delta}(\,\wt{c}\,)} \right)_{\sodkn}
\en
where $\Delta= r(a)+r(b)-r(c)$, and $r(\tau)$ denotes the number of cells
in the tableau $\tau$.
Using this in (\ref{quotmult}) gives
\eq
    \left( {M_{\vphantom{\wt{b\mathstrut}}a b}}^{c} \right)_{\cR_{n,k}}=
  \left(  {M_{\, \wt{a\mathstrut}\, \wt{b\mathstrut} }}^{\;\, \wt{c}} \,
       \right)_{\cR_{k,n}}  \; .
\label{rldual}
\en
The equalities $\widetilde{\sigma(c)}=\widetilde{c}$,
$\Delta^{ab}_{\sigma(c)}=\Delta^{ab}_c + 1$ (mod $2$),
and $\sigma^2 =1$ have been used here.
Identity (\ref{rldual}) is just the statement
that $\cfT_{n,k}/\cC_k$ and $\cfT_{k,n}/\cC_n$ are isomorphic
as rings under tableau transposition.

\subtopic{The Transposition Quotient is the Cominimal Quotient}

Now we shall show that the rings $\cR_{n,k}$ (\ref{quotring})
are identical to the cominimal quotients of the
$\cfT_{n,k}$ just defined, \ie\ that
\eq
          \cR_{n,k} = {\cfT_{n,k} \over \cC_k}
\label{isotoget}
\en
under the correspondence between ring elements
given by the natural correspondence of tableau labels:
\eq
              \phi_{\bsub{a}} \leftrightarrow \char_{\bsub{a}}  \; .
\label{tabcorr}
\en

Since $\phi_i \leftrightarrow \hchi_i$, comparison of (\ref{quotring})
and (\ref{comquot}) shows that we must demonstrate the equivalence
\eq
\cI_k \equiv    \cF_k \cdot \cC_k
\label{Fring}
\en
under the correspondence (\ref{tabcorr}).

The relations that generate the ideals $\cI_k$ (\ref{genvan})
are transparently special cases of the cominimal equivalence
relations (\ref{comin}) if $n \leq k+1$.  In general they
also include examples of the fusion identities implied by the extended
Speiser algorithm (this is shown in part three of the appendix).
In all cases the relations that
generate $\cF_k \cdot \cC_k$ imply the relations that generate
the transposition ideals $\cI_k$. Therefore $\cI_k$ is a subideal
of $\cF_k \cdot \cC_k$.

In the first two parts of the appendix we use
the determinant formula (\ref{basdet}),
and the extended Speiser algorithm to show the converse, namely, that
the generators (\ref{genvan}) imply the entire set of
cominimal equivalence relations (\ref{comin}) and the
entire set of fusion relations (\ref{fusrule}). Therefore
$\cF_k \cdot \cC_k$ is a subideal of $\cI_k$, and
the equivalence (\ref{Fring}) follows.

{}From (\ref{rldual}) it follows that the rings
related by interchange of rank and reduced level are isomorphic:
\eq
   \cR_{n,k} \equiv \cR_{k,n} \; ,
\label{Riso}
\en
with the isomorphism given by tableau transposition.

It is well known that there is a solution for basis elements
$\phi_a$ of the $\sodnk$ fusion ring for
every integrable highest weight representation $r$ of $\sodnk$ of the
form
\eq
         \phi_a(r)  = {S_{ar}\over S_{0r}} \; .
\label{solution}
\en
Given this fact and isomorphism (\ref{isotoget})
it follows that the solutions of the
$n$ polynomial equations (\ref{genvan}) for the $n$ variables $\hchi_i$ are
exactly those solutions (\ref{solution})
of the fusion rule algebra that in
addition satisfy the cominimal equivalence relations (\ref{comin}).
Since
\eq
       S_{\sigma(a) r} = \cases{ +S_{a r} & for $r$ a tensor \cr
                                 -S_{a r} & for $r$ a spinor $\;$,}
\label{spinrm}
\en
imposition of the cominimal equivalence relations (\ref{comin})
excludes precisely the solutions (\ref{solution}) for which $r$ is
a spinor.  Therefore, the values of the fundamental characters
$$
              \hchi_i = {S_{it}\over S_{0t}} \; ,
$$
where $t$ is any tensor representation
specified by a tableau with $\row_1 \leq k$,
give a complete set of solutions of (\ref{genvan}).
{}From this result and determinant formula (\ref{basdet}),
the values of the $\cR_{n,k}$ basis character
corresponding to an arbitrary $n$ row tableau $a$ with
$\row_1(a) \leq k$  are
\eq
         \char_{\bsub{a}}(t)   =  {S_{at}\over S_{0t}} \; ,
\en
where $t$ (a tableau with $\row_1 \leq k$) labels the solution.

It follows from (\ref{spinrm}) that the
relation between the $\cR_{n,k}$ multiplicities (\ref{quotmult}) and
the $\sodnk$ modular matrix elements,
\eq
      {M_{ab}}^c = 4 \sum_{\hbox{\bsub{{t \atop \row_1(t)\leq k}}} }
                          {S_{at} S_{bt} S_{ct} \over S_{0t}} \; ,
\en
involves a sum only over the integrable {\em tensor} representations
with $\row_1 \leq k$.
The matrices $\cS_{ab}$ that diagonalize the $\cR_{n,k}$ tableau basis
are related to the $\sodnk$ modular transformation matrices
$S_{ab}$ (note the difference in typeface) by
\eq
       \cS_{ab} = 2 S_{ab}  \; .
\label{SforR}
\en
The orthonormality condition
\eq
       \sum_{\hbox{\bsub{{t \atop \row_1(t)\leq k}}} }
                     \cS_{at} \cS_{tb} = \delta_{ab}
\label{Snorm}
\en
for the matrices $\cS_{ab}$ then follows from
the orthonormality condition for $S_{ab}$.

\topic{Fusion Rings and Topological Metrics}

After pointing out that the $\cR_{n,k}$ are fusion rings,
but that they are not the fusion rings of any conformal field theory,
we identify a special field whose properties ensure that
the $\cR_{n,k}$ will admit an invertible topological metric.

\subtopic{The Quotients $\cR_{n,k}$ as Fusion Rings}

A rational fusion ring is a finite-dimensional,
commutative, associative, $\ZZ$-ring
with identity that also has an involutive conjugation automorphism
and a special {\em multiplicity basis}
in which the structure
constants ${N_{ab}}^{c}$ that define the fusion product (\ref{fusrule})
between the special basis elements are non-negative integers.\cite{ffrev}
The rings $\cR_{n,k}$ clearly have all these properties
(all fields are self-conjugate and the
tableau basis is the special multiplicity basis), so that they
are rational fusion rings.

We will now show that the $\cR_{n,k}$ are not, in general,
the fusion rings of {\em any} conformal field  theory.
This result is expected since (diagonal) modular invariant
partition functions of the
$\sodnk$ fusion ring (of conformal scalars)
with the spinors removed are not known.
While the cominimal quotient of the entire $\sodnk$ fusion ring likely
is the fusion ring of a conformal field theory, the presence of the
\setcounter{footnote}{1}
spinors makes finding potentials that
might lead to integrable $\cN=2$ theories
more difficult.

The simple case $\cR_{1,1}$, which is associated with
$\hbox{\sc so}(3)_3$, has two elements ${\bf 1}$, and \fund~, which
obey the fusion rule
\eq
  \fund\; \star\; \fund  =  {\bf 1} + 2\; \fund \; .
\en
It is known that this fusion ring cannot satisfy certain
constraints on the conformal weights and modular matrix elements
required in a conformal field theory.\cite{badrule,ffrev}

To see how this works out in a more complicated case
consider the ring
$\cR_{2,1}$, which is associated with
$\hbox{\sc so}(5)_3$. Its three elements correspond
to the tableaux ${\bf 1}$, \fund~, and \yfund1~.
If this ring were the fusion ring of a conformal field theory, then
the properly normalized
matrices $\cS$ and $\cT$ that diagonalize the
fusion rules would have to satisfy
(since $\cS$ is real and all fields are self-conjugate)
\eq
	    (\cS \cT)^3 = I   \; .
 \label{cond1}
\en
In addition, the conformal weights $h_i$ and the symmetric fusion
coefficients $N_{ijk}$ must satisfy Vafa's constraint\cite{vafaconst}
\eq
   \prod_r (\alpha_{i} \alpha_{j} \alpha_{k} \alpha_{l})^{N_{ijr} N_{rkl}}
= \prod_{r} \alpha_{r}^{N_{ijr} N_{rkl} + N_{ikr} N_{rjl} + N_{ilr} N_{rjk}},
 \label{cond2}
\en
where $\alpha_{i} = e^{2 \pi i h_{i}}$.

The nontrivial fusion rules of $\cR_{2,1}$ are
\eq
    \begin{array}{rcl}
           \fund\; \star\; \fund & = & {\bf 1} + \fund + \yfund1 \\[0.3cm]
          \fund\; \star\; \yfund1 & = & \fund + 2\; \yfund1 \\[0.3cm]
          \yfund1\; \star\; \yfund1 & = & {\bf 1} + 2\; \fund + 2\; \yfund1
     \end{array}
\label{fusexam}
\en
and the diagonalizing matrix $\cS_{ab}$ is
\setlength{\unitlength}{0.13cm}
\eq
          \cS_{ab} = {1\over 2\sqrt{3}} \times
\bordermatrix{
       & {\bf 1}    &  \lphm   \fund        &  \lph \yfund{1.1} \rph \cr
{\bf 1}& \sqrt{3} - 1  &   \lphm  2   & \sqrt{3}  +1 \cr
\fund  &   2 &      \lphm    2   &      - 2     \cr
\yfund{1.1}
       & \sqrt{3} + 1 &    -2         &   \sqrt{3} - 1   }\; .
\en

The $15$ equations (\ref{cond2}) yield three independent
constraints on the $\alpha_i$, which are,
\setlength{\unitlength}{0.13cm} with the indicies $(i,j,k,l)$ of
(\ref{cond2}) listed on the right,
\eq
     \begin{array}{rclr}
      &   & \phantom{1_{\setlength{\unitlength}{0.13cm}\, \yfund{1.5}}}
       \qquad (\ijkl{i}\, ,\ijkl{j}\, ,\ijkl{k}\, ,\ijkl{l})\\[0.2cm]
\alpha_1 & = & 1_{\phantom{\setlength{\unitlength}{0.13cm}\, \yfund{1.5}} }
\setlength{\unitlength}{0.2cm}%
       \qquad (\bfid\, ,\bfid\, ,\bfid\, ,\bfid)\\[0.2cm]
\setlength{\unitlength}{0.13cm}%
\alpha_{\, \fund}^9 & = &
       \alpha_{\, \yfund{1.5}}^3
\setlength{\unitlength}{0.2cm}%
       \qquad(\fund\, , \fund\, , \fund\, ,\fund)\\[0.3cm]
\setlength{\unitlength}{0.13cm}%
\alpha_{\, \fund}^6 & = & \alpha_{\, \yfund{1.5}}^3
          \setlength{\unitlength}{0.2cm}%
         \qquad(\fund\, ,\fund\, ,\fund\, ,\yfund{0.7}\;) \; .
\end{array}
\en
It follows that the conformal
\setlength{\unitlength}{0.13cm}
weight of the identity is integral ($\alpha_{1} = 1$), and that
there are nine possible solutions to the two remaining constraints
$\alpha_{\fund}^{3} = 1$ and $\alpha_{\yfund{1.1}}^{3} = 1$.

For any Virasoro central charge $c$,
\eq
     \cT_{i} = e ^ {[2 \pi i (h_{i} - c/24)]} \; ,
\en
and the identity-identity matrix element of (\ref{cond1}) becomes
\eq
\sum_{i,j} \cS_{0i}\cS_{ij}\cS_{j0} \alpha_{i} \alpha_{j} = e^{\pi i c /4} \; .
\label{spcond}
\en
By explicit computation, we find that none of
the nine possible values for the $\alpha_{i}$ satisfy (\ref{spcond})
for any real central charge $c$.
Therefore, $\cR_{2,1}$
cannot be the fusion ring of {\em any} conformal field theory.
While it is difficult to adapt this form of argument to the
arbitrary case, we expect that the rings $\cR_{n,k}$ do not in general underlie
conformal field theories. Although this might be expected (since the
spinors have been omitted) it raises the interesting problem of
characterizing the conditions under which a fusion quotient of a
conformal-field-theory fusion ring also underlies a conformal field theory.

\subtopic{A Candidate Topological Metric}

The topological Landau-Ginzburg theories described in section five have
chiral rings isomorphic to the fusion rings $\cR_{n,k}$. In the
natural tableau basis for these rings the topological
metric $\eta_{ab}$ is given in
terms of the $\cR_{n,k}$ multiplicities (\ref{quotmult}) by
\eq
         \eta_{ab} = {M_{ab}}^{\bbox} \; ,
\label{etadef}
\en
where $\bbox$ is the tableau with $n$ rows each of length $k$.
It can be seen that $\eta_{ab}$ is an invertible matrix as follows.

Let $\rho$ denote the the operation of tableau complement
in an $n\times k$ rectangle. In terms of the row
lengths, $\row_i(a)$, of a tableau $a$, the tableau $\rho(a)$ has row
lengths $\row_i(\rho(a))= k - \row_{n+1-i}(a)$. Diagrammatically
the operation $\rho$ forms the complement of the tableau
$a$ in the $n\times k$ rectangle, and then rotates this complement
by $180$ degrees to put it in standard position. The
complement of the identity, $\rho(0) = \bbox$, is the unique tableau with
the maximal number of cells ($nk$).

In order to show that $\eta_{ab}$ is invertible, we will only need to
calculate ${M_{ab}}^{\bbox}$ when
$r(a)+r(b)\leq nk$. We will perform the calculation by
imposing the rank and fusion modification rules on
the tableau product (\ref{mixal}) of $a$ and $b$.
All the terms in (\ref{mixal}) with $d$ not equal to the identity ($0$)
have fewer boxes than $\bbox$. This means that only the terms with
$d=0$, which are given by applying the Littlewood-Richardson
rule to $a$ and $b$, can directly produce $\bbox$.
All we need to know about the modification rules is the
\begin{quote}
\noindent {\it Modification Rule Property:}
Any sequence of modification rules that connect a
tableau with cells outside the $n\times k$ rectangle
(\ie, one with $\row_1 > k$ or $\col_1> n$) to one
inside this rectangle (\ie, one with $\row_1\leq k$ and
$\col_1 \leq n$) relates the outer tableau
to an inner tableau with {\em fewer} cells.
\end{quote}
\eq
\label{modprop}
\en
If $r(a) + r(b) < nk=r(\bbox)$ then (\ref{modprop})
implies that $\bbox$ cannot possibly appear in (\ref{mixal}) so that
$\Mabhat=0$ for all such cases.
In the cases with $r(a) + r(b) = nk$, $\Mabhat$
just equals, again due to (\ref{modprop}), the Littlewood-Richardson
multiplicity, $\Labhat$.
In the context of $\sunless$
representation theory,
the tableau $\bbox$ just denotes the $\sunless$
identity representation, and $\rho(a)$, the complement of $a$ in an
$n\times k$ rectangle, denotes the complex conjugate representation,
$\overline{a}$, of the $\sunless$ representation $a$. Since $\Labhat$
equals the multiplicity of the identity in
the $\sunless$ tensor product $a\otimes b$,
since $a\otimes b$ does not
contain the identity unless $b=\overline{a}$,
and since $a \otimes \overline{a}$ contains the identity just once,
it is clear that $\Labhat =\delta_{b\rho(a)}$ for $n\geq 2$.
(The same result holds for $n=1$ trivially.)
Therefore,
\eq
      \eta_{ab} = \delta_{b\rho(a)}    \qquad {\rm for}
   \quad r(a) + r(b)\leq nk \; .
\label{neededfact}
\en
Unlike $\Labhat$, the fusion multiplicities
$\Mabhat$ do not vanish when $r(a)+r(b)>nk$.
Since each row of $\eta_{ab}$ in fact contains several nonzero elements,
demonstration of its invertibility requires a little more work.

Order the basis tableaux
that label the rows $a$ and columns $b$ of $\eta_{ab}$
in terms of the increasing number of cells $r(a)$ and $r(b)$ in the
respective basis tableaux.
This means that any tableaux $a$ ($b$) with more cells appears
below (to the right of) tableaux with fewer cells. For example,
the identity $0$ labels the first row and column and $\rho(0)$
\setlength{\unitlength}{0.2cm}
labels the last row and column.  Similarly, $\fund$
labels the second row and column, and $\rho(\fund)$ labels the
next-to-last  row and column.
Let
$$
   D =  {n+k \choose n}
$$
denote the dimension of the matrix $\eta_{ab}$. This is just the
dimension of the ring $\cR_{n,k}$ (\ie, the
number of $n$ column tableaux with $\row_1 \leq k$).
{}From (\ref{neededfact}) we
see that any matrix element with row label $a$
(and appearing at the $i^{\rm th}$ row), and with column label $b$ with
$r(b) < nk-r(a)$ (so that it appears in the
$j^{\rm th}$ row with $j< D+1-i$) vanishes.
In the remaining cases, there occur subblocks positioned
along the $(i,D+1-i)$ {\it anti-diagonal} with each
subblock labeled by tableaux with the {\em same} number of cells.
Result (\ref{neededfact}) implies that in each row of any subblock there
is only one nonzero entry (which is unity)
so that a rearrangement of columns will, in all cases, put $\eta_{ab}$
in the form of a matrix with all
anti-diagonal matrix elements $(i,D+1-i)$ equal to
unity, and all upper anti-triangular matrix elements
(\ie, all those having coordinates $(i,j)$ with $j< D+1-i$) equal to
zero. The determinant of such a ($D\times D$) matrix is
$(-1)^{D(D-1)/2}$ so that
\eq
        |\det  \eta_{ab}| = 1
\label{detres}
\en
and the matrix inverse of $\eta_{ab}$ exists.
It is clear from the cofactor expression for $\eta^{-1}_{ab}$
and the integrality of the determinant that the inverse of
the integral matrix $\eta_{ab}$ is also an integral matrix.

\setlength{\unitlength}{0.18cm}
For example, the complete matrix $\eta_{ab}$
for $\cR_{2,1}$, with the basis ordering ${\bf 1}$, \fund~, \yfund{1.1}~
and with $\bbox = \yfund{1.1}$ , can be written down directly from the
fusion rules (\ref{fusexam}):
\eq
 \eta_{ab}= \bordermatrix{   &  {\bf 1} & \fund &  \yfund{1.2} \cr
                           {\bf 1} &      0 & 0 & 1 \cr
                           \fund   &    0 & 1 & 2 \cr
                         \yfund{1.1} &    1 & 2 & 2 }
   \qquad {\rm and} \qquad
 \eta^{-1}_{ab}= \bordermatrix{
                   &  \lphm {\bf 1} &\lphm \fund &\lphm  \yfund{1.2} \cr
             {\bf 1} & \lphm  2 & -2 &\lphm 1 \cr
             \fund   &    -2 & \lphm 1 & \lphm 0 \cr
        \yfund{1.1} & \lphm   1 & \lphm 0 &\lphm 0 }  \; .
\en
As claimed, all entries in $\eta^{-1}$ are integers.

The complement operation also figures prominently in
the $\sunk$ and $\spnk$ deformations of the Grassmannian
superpotentials.  In these cases the complement
map is a simple-current fusion-rule automorphism, and the
invertibility of the topological metric follows directly from
this fact. In the present $\sonk$ case, the complement operation
retains the property of leading to an invertible
topological metric, even though it is not a fusion rule automorphism.
In the following subsection we recall that the complement
operation is related to Poincar\'e duality.

\subtopic{The Grassmannian Schubert Calculus}

While the results in this section are (presumably well-known but
somewhat) implicit in previous work,\cite{gh,gepner,intri}
we state them here to make salient certain relevant points.

The integral homology of the Grassmannian
\eq
            \cG_{n,k}=  {U(n+k)\over U(n) \times U(k)}
\en
is freely generated by the Schubert cycles (or subvarieties) $\sigma_a$
where $a$ is a tableau with first column length $m_1(a)\leq n$
and first row length $\row_1(a)\leq k$. All cycles with
$m_1>n$ or $\row_1 > k$ are null.\cite{gh}
Since the intersection of two such subvarieties can be written as
an integral linear combination of the generating Schubert cycles,
\eq
           \sigma_a \star \sigma_b = \sum_c {G_{ab}}^{c} \sigma_c ,
\label{cohop}
\en
the Schubert cycles form a $\ZZ$-ring under the intersection product $\star$.
With $K^{\sigma}_i$ for $i=1,\ldots, k$ denoting the
cycles corresponding to single-row tableaux of length $\row_1=i$, an
arbitrary cycle can be written as a polynomial
in the $K^{\sigma}_i$ via the $\col_1(a)\times \col_1(a)$
determinant\cite{gh}
\eq
           \sigma_a = \det | K^{\sigma}_{\row_{i} -i+j}| \; .
\label{fgrassdet}
\en
Here, $K^{\sigma}_{i}$ with $i >k$ or $i<0$ should be set to zero.
Since this determinant formula is identical to the expansion of
$U(n)$ characters in terms of row characters of covariant
$U(n)$ tableaux, the tableau product given by polynomial
multiplication ({\em without} setting $K^{\sigma}_i$ to zero for $i>k$)
is just that given by the Littlewood-Richardson
rule. To recover the actual intersection product we
must impose the
\begin{quote}
{\it Grassmannian Modification Rules}:
Remove any tableau with first column length greater than $n$
(the rank modification rule)
and any tableau with first row length greater than $k$
(the fusion modification rule).
\end{quote}
Note that these modification rules possess property (\ref{modprop}).
If just the rank modification rule is imposed, the ring we obtain is
exactly the ring of $U(n)$ characters, \ie, the $U(n)$ representation
ring of covariant tensors.  This means that the
intersection ring of Schubert cycles is a quotient of the
$U(n)$ representation ring of covariant tensors by the ideal
defined by the fusion modification rule.
Since the intersection multiplicities ${G_{ab}}^{c}$ can be
calculated by imposing the Grassmannian
rank and fusion modification rules on the
Littlewood-Richardson product, the proof of the last section applies
directly so that
\eq
{G_{ab}}^{\bbox}=\delta_{b,\rho(a)} \; .
\label{grassneededfact}
\en
In the context of algebraic geometry this result is demonstrated
by an appeal to Poincar\'e duality (and the fact that analytic
cycles, such as the Schubert cycles considered here, intersect
positively).\cite{gh}

{}From the manifest symmetry under interchange of $n$ and $k$ given by
row and column interchange, one also has
\eq
     \sigma_a = \det | \hchi^{\sigma}_{\col_{i} - i+j}|
\label{grassdual}
\en
where $\hchi^{\sigma}_i=0$ if $i>n$ or $i<0$.
{}From this determinant it follows that the conditions
\eq
         \hchi^{\sigma}_{i} = 0
              \qquad {\rm for}\quad i=n+1, \ldots
\en
generate the ideal defined by the rank modification rules.
Then, it follows from (\ref{fgrassdet}) that the fusion ideal is
generated by the polynomial identities
\eq
      K^{\sigma}_{i}(\hchi^{\sigma}_j) = 0
\qquad {\rm for} \quad i=k+1, \ldots, k+n \; .
\label{grassvan}
\en
These polynomials integrate to a quasi-homogeneous potential.
The $N=2$ superconformal field
theories characterized by these superpotentials\cite{lerchewarn}
have the intersection rings (\ref{cohop}) as chiral rings.
Since the topological metric
is given by $\eta^0_{ab} = {G_{ab}}^{\bbox}$, (\ref{grassneededfact})
yields the well-known result
\eq
        \eta^0_{ab}  = \delta_{b\rho(a)} \; ,
\label{grassmet}
\en
and the invertiblity of the
Grassmannian topological metric follows immediately.

\topic{Potentials with the $\cR_{n,k}$ Fusion Rings as Local Rings}

Now we are ready to construct potentials that have the rings
$\cR_{n,k}$ as local rings, \ie\ we will find a $\cV$ such that
\eq
        \cR_{n,k} =
              {\ZZ [ \hchi_{1}, \ldots, \hchi_{n} ]\over {\rm d}\cV }
\en
for each $n$ and $k$.  The construction is natural in the context of
$\son$ group theory. Although many potentials that have $\cR_{n,k}$
as a local ring can be found,\cite{aharoni}  we claim that those constructed
here are special, in that they do yield (we expect integrable)
$\cN=2$ Landau-Ginzburg theories, and in that
the associated (twisted $\cN=2$) topological theories have several
special properties, as we shall see.
(We have found that the covering-space potentials of ref.~\dcite{cresc}
do not always project onto potentials in the physical variables,
since, at least
in several examples, {\em all\/} the critical-point vanishings
appear in the Jacobian rather than in the derivatives of
the physical potential.)

The characters $\hchi_{0}, \hchi_{1}, \ldots$ can be written
as specialized versions of the elementary symmetric functions
\eq
  \hchi_j = E_{j} = E_j(q_1, \ldots, q_n,q_1^{-1},\ldots,q_n^{-1},1)
\label{charsym}
\en
where the $E_{j}$ are defined in terms of
the auxiliary variables
\setcounter{footnote}{1}
$q_{i}$ ($i=1,\ldots,n$)\footnote{Without imposition of the $\cI_k$ generating
relations, the $q_i$ are exponentials of certain Cartan subalgebra elements;
when the the relations (\ref{genvan}) are imposed they take fixed
values (\ref{auxcrit}).}
via the generating function\cite{fh}
\eq
    E(t) = \sum_{j=0}^{\infty} E_{j} ~t^{j} =
   (1+t)  \prod_{i=1}^{n} (1+ q_{i} t) (1 + q_{i}^{-1}t)   ~~.
\label{symdef}
\en
It follows that $E_{0}=1$,
$E_{2n+1-j} = E_{j}$ for $j=0,\ldots,2n+1$, and $E_{j} =0$ if $j> 2n+1$,
so that the identities (\ref{rmodgen}) are indeed satisfied.

It is a standard result\cite{fh} that the determinant
(\ref{basdet}) and the identity (\ref{charsym}) imply that
the character $K_i$ of the single row tableau of length $i$
satisfies
\eq
K_{i} = H_i - H_{i-2} \quad,
\label{rowsym}
\en
where  the $i^{\rm th}$ complete
symmetric function $H_i$ of the variables $q_{i}$ can be obtained
from the generating function\cite{fh}
\eq
    H(t) = \sum_{j=0}^{\infty} H_{j} ~t^{j} = {1 \over 1- t}
     \prod_{i=1}^{n} {1 \over (1-q_{i} t) (1- q_{i}^{-1}t)}  ~~.
\label{comsymdef}
\en
Note that
\eq
         H(t) E(-t) = 1  \; .
\label{pointout}
\en

The similarity between identities (\ref{symdef}--\ref{comsymdef})
and those
used in the construction of the $\spnk$ potentials in ref.~{\dcite{brsym}}
suggests that we consider the generating function
\eq
\begin{array}{rcl}
   V(t) & = & \log E(t) \\[0.2cm]
  & = & \ds \log (1 + t) \prod_{i=1}^n (1 + q_{i} t)(1 + q_{i}^{-1} t) \\
   & = & \ds  \sum_{m=1}^{\infty} (-1)^{m-1} V_{m} t^{m} \; ,
 \end{array}
\label{auxpotf}
\en
in which case
\eq
    V_{m} = {1 \over m} \sum_{i=1}^{n} (q_{i}^{m} + q_{i}^{-m})  +
{1\over m}   \; .
\label{auxpot}
\en
Differentiation of (\ref{auxpotf}) and
the use of (\ref{pointout}) yields
\eq
     {\partial V_{m} \over \partial E_{i}} =
     \sum_{{\sst 0 \leq j \leq 2n+1}} (-1)^{j+1} H_{m-j}
    {\partial E_{j} \over \partial E_{i}}  \; .
\en
Since $\hchi_j = E_{j} = E_{2n+1-j}$, we find
\eq
     {\partial V_{m} \over \partial \hchi_{i}} =
   (-1)^{i+1} (H_{m-i} - H_{m-2n+1-i} )  \qbox{for} 1 \leq i \leq n \; .
  \label{dervan}
\en
We now consider, as candidates for the potentials we want,
the differences
\eq
     \cV_{m} = V_{m} - V_{m-2} \; ,
\label{pot}
\en
which satisfy
\eq
     {\partial \cV_{m} \over \partial \hchi_i} =
   (-1)^{i+1} (K_{m-i} - K_{m-2n-1+i} )  \qbox{for} 1 \leq i \leq n \; .
  \label{realvan}
\en
Although the potentials $\cV_m$ are defined in terms of
the $q_i$, it is clear that they are also
polynomials in the $\hchi_{i}$.
Upon setting $m=n+k+1$, we see that the critical point conditions
$$
{\partial \cV_{n+k+1}\over \partial \hchi_i} = 0 \rlap{$\qquad 1\leq i \leq n$}
$$
exactly coincide (by setting $i=n+1-j$ in eq.~\ref{realvan}) with the
generators of the transposition ideal $\cI_k$ (\ref{genvan}).

Therefore, the local ring of the polynomial
$\cV_{n+k+1}$, written in terms
of the fundamental variables $\hchi_{1},\ldots,\hchi_{n}$,
is, in a natural way, exactly $\cR_{n,k}$.

While the shortest path to obtaining $\cV_{n+k+1}$ as a polynomial
in the $\hchi_i$, for given $n$ and $k$, is
to integrate (\ref{realvan}), it is useful to have in hand
a recursion relation for the potentials.
Since the coefficient of $q^j$ in the expansion of
the polynomial
\eq
    P(q) =  \prod_{i=1}^{n} (q-q_{i})(q-q_{i}^{-1})
 \label{qpoly}
\en
is given by
$$
\sum_{p=0}^{j} (-1)^{p} \hchi_{p}  \; ,
$$
and since the $q_i^{\pm 1}$ all satisfy $P(q)=0$,  one
obtains from this identity the recursion relation
\eq
   \sum_{i=0}^{2n-1} \left( \sum_{j=0}^{i} (-1)^{j} \hchi_{j}  \right)
[ (l+2n-i) V_{l+2n-i}-1] +
   \left\{  \begin{array}{cl}
     2n & l=0  \\
   l V_{l}-1 & l>0
       \end{array}
   \right\}   = 0   \; .
 \label{recur}
\en
for the $V_m$.

As an example, the recursion
relation for $n=2$ yields the potentials related to
$\hbox{\sc so}(5)_{2k+1}$. With
$x=\hchi_{1}$ and $y=\hchi_{2}$ it reads,
in terms of the auxiliary quantity
$U_m = m V_m -1$,
\eq
    U_{l+4} - (x-1) U_{l+3} + (y-x+1) U_{l+2} - (x-1) U_{l+1} +
\left\{  \begin{array}{c}
      ~4~~~~l=0  \\
     U_{l} ~~l>0
       \end{array}
   \right\}   = 0
 \label{rec-so5}
\en
where $U_{1}= x-1$, $U_{2}= x^2 - 2y - 1$, and
$U_{3}= x^3 - 3xy + 3y - 1$ are the proper
initial conditions. Then, $\cV_m = U_m/m - U_{m-2}/(m-2)-2/(m^2-2m)$
gives the actual potential.

\topic{$\cN=2$ Landau-Ginzburg Theories}

Consider the $N=2$ supersymmetric theory characterized by the
superpotential $W_{n,k} = \lambda^{n+k+1} \cV_{n+k+1}(\Phi_{i})$
for $n$ chiral superfields $\Phi_i$.
A finite number of states are topological (in that their correlation
functions do not depend on the locations of the fields),
and form a closed ring\cite{martharv,lerchewarn}
\eq
      {\CC[ \Phi_{1},\ldots,\Phi_{n} ] \over {\rm d}W_{n,k}}
\en
clearly isomorphic to $\cR_{n,k}$. A tableau basis for this chiral ring
is given by associating with any tableau $a$ with $\row_1 \leq k$
the field
\eq
   \psi_a = {\lambda^{r(a)} \over 2}
            \det | \Phi_{\col_i-i+j} + \Phi_{\col_i-i+2-j}|
\label{phidet}
\en
where $\Phi_i=0$ for $i<0$. We take the analog of (\ref{rmodgen})
\eq
  \begin{array}{ll}
 \Phi_{n+j} = \Phi_{n-j+1} & \rlap{\qquad $1\leq j \leq n+1$} \\
  \Phi_{n+j} = 0            &  \qquad \rlap{\qquad $j> n+1 $}
          \end{array}
\label{phidef}
\en
as a definition of $\Phi_i$ for $i>n$. While this form of the potential
and chiral ring basis
makes the isomorphism of the chiral ring with $\cR_{n,k}$
transparent, the field redefinition $\Phi_i \rightarrow \lambda^{-i} \Phi_i$
makes the r\^ole of the deformation parameter $\lambda$ clear.
Under this rescaling substitution, the
critical point vanishing conditions become
\eq
{\partial W_{n,k}^{\lambda}(\Phi_j) \over \partial \Phi_i} =
 (-1)^{i+1} (K_{k+n+1-i}^{\lambda}(\Phi_j) -
      \lambda^{2n+1-2i} K_{k-n+i}^{\lambda}(\Phi_j)) = 0
\label{phivan}
\en
where we have set $W_{n,k}^{\lambda}(\Phi_j)= W_{n,k}(\lambda^{-j}\Phi_j)$,
and where, from (\ref{phidet}),
\eq
   \begin{array}{lcl}
   K_j^{\lambda}(\Phi_i) & = &
     \half \lambda^{j}
            \det | \lambda^{-(\col_p-p+q)}\Phi_{\col_p-p+q} +
                   \lambda^{-(\col_p-p+2-q)}\Phi_{\col_p-p+2-q}| \\[0.2cm]
    & = & \half \det | \Phi_{\col_p-p+q} + \lambda^{2q-2} \Phi_{\col_p-p+2-q}|
   \end{array}
\en
with $p,q=1,...,j$.
The column lengths are all one ($\col_p=1$), but we have left this
implicit for clarity.
The limit $\lambda \rightarrow 0$ of (\ref{phivan}) gives
\eq
{\partial W_{n,k}^{0}(\Phi_j) \over \partial \Phi_i} =
 (-1)^{i+1} \det| \Phi_{m_p-p+q}| = 0
\en
which are exactly the vanishing conditions (\ref{grassvan}) for the
cohomology ring of the
Grassmannian $\cG_{n,k}$ which integrate to the known
quasi-homogeneous, Grassmannian superpotentials.
Similarly, under this scaling the tableau basis (\ref{phidet}) becomes
$$
\psi_a = \half \det | \Phi_{\col_p-p+q} + \lambda^{2q-2} \Phi_{\col_p-p+2-q}|
$$
so that, as $\lambda$ goes to zero, it flows directly to the standard
tableau basis of the Grassmannian chiral ring (\ref{grassdual}).
In particular, the special field $\psi_{\bbox}$ flows smoothly
to the unique field with maximal charge.
The \LG potentials of these $\sodnk$ related fusion rings
therefore arise as deformations of the same
$\cN=2$ superconformal field theories that can be deformed
to obtain the $\spsnk$ and $\sunmorek$ fusion potentials.
It would be interesting if all three of these deformation
directions yield integrable theories of the kinks that
interpolate between the critical points.

One can, of course, obtain the same result by scaling the $q_i$
in section four by $\lambda^{-1}$, in which case one sees that
the defining form of the potential,
\eq
     (W_{n,k})_{\sst {\rm quasi-hom}} =
    { 1 \over n+k+1} \sum_{i=1}^{n} q_{i}^{n+k+1} \; ,
\en
is identical to that expected for the Grassmannian $\cG_{n,k}$
in terms of the variables $q_i$.

While the tableau basis (\ref{phidet}) is a natural deformation
of the Grassmannian chiral ring, the metric $\eta_{ab}$, although invertible,
is not anti-diagonal, and so depends on the parameter $\lambda$.
Therefore the tableau basis is not flat and it is
unclear how to establish a general connection to the
canonical bases that arise in conformal perturbation theory.

Since the critical points of the rings $\cR_{n,k}$
are completely non-degenerate
we are dealing with completely massive theories for non-zero $\lambda$.

In the $n=1$ case, the one variable superpotential
$W_{1,k}(\Phi)$ with local ring  $\cR_{1,k}$
(necessarily) gives a deformation of an $\cN=2$ minimal models.
For example, the first two potentials are,
\eq
   \begin{array}{rcl}
   W_{1,2}^{\lambda}(\Phi) & = &
  {\tx {1\over 4}} \Phi^4 - \lambda \Phi^3 + 2\lambda^3 \Phi \\[0.2cm]
  W_{1,3}^{\lambda}(\Phi) & = &
 {\tx {1\over 5}} \Phi^5 - \lambda \Phi^4 + {\tx {2\over 3}}\lambda^2 \Phi^3
                 + 2\lambda^3 \Phi^2 - \lambda^4 \Phi - \lambda^5
   \end{array}
\en
In order to compare these with the known integrable
deformations\cite{intzero} the field $\Phi$ must
first be shifted to remove the leading term of the deformation.
In general, the deformation in this new basis is a
linear combination of the $\Phi^j$ for $j=1, \ldots, n+k-1$ so
that they appear to be different massive theories than those
considered previously.

The first interesting multi-variable potential
has $\cR_{2,1}$ as a chiral ring. With $x=\Phi_1$ and
$y=\Phi_2$ it is
\eq
\lambda^{-4}W_{2,1}(x,y) = {\sct {1 \over 4}} x^{4} + \half y^{2} - x^{2} y
+ x y - \half x^{2} + y - x   \; .
\en
For comparison, the two-variable potentials
related to $\Sp(2)_1$ and $\hbox{\sc su}(3)_1$, are
\eq
 \begin{array}{lcl}
\lambda^{-4}W_{\Sp(2)_{1}}(x,y) & = & {\sct {1 \over 4}} x^{4} +
\half y^{2} - x^{2} y + y - \half  \\[0.2cm]
\lambda^{-4} W_{\hbox{\sc su}(3)_1}(x,y) & = &
{\sct {1 \over 4}} x^{4} + \half y^{2} - x^{2} y + x \; .
\end{array}
\en
All three have the same leading terms
\eq
{\sct {1 \over 4}} x^{4} + \half y^{2} - x^{2} y
\en
which is the quasihomogeneous potential
whose local ring is the cohomology
ring of the Grassmannian $U(3)/U(2) \otimes U(1)$.

While we will leave the quesion of the quantum
integrability of the theories based on these potentials for another
work, we will now show that the $\cN=2$ topological theories that can
be constructed from these potentials enjoy rather special properties.

\subtopic{Twisted $\cN=2$ Topological Field Theories}

By twisting the energy-momentum tensor with the $U(1)$ generator
present in any $\cN=2$ theory, the topological character of
the chiral rings $\cR_{n,k}$
can be exploited to produce $\cN=2$ {\em topological}
field theories composed entirely of the
chiral primary fields.\cite{twistop}
Due to the close connection with the
$\sodnk$ fusion algebra the topological
field theories can be completely solved.

The genus $g$ correlation functions with an insertion of
an arbitrary function
$f$ of chiral superfields $\Phi_{i}$, with superpotential $W$, is\cite{vafa}
\eq
      \vev{f(\Phi_{i})}_{g} = \sum_{{\rm d}W=0} H^{g-1} f(\Phi_{i})
\label{residue}
\en
where the handle operator\cite{witten} is\cite{vafa} (using the
normalization of ref.~\dcite{intri})
\eq
     H = (-1)^{n(n-1)/2} \det (\partial_{i} \partial_{j} W) ~~.
\label{vafone}
\en
The sum is over the critical points at which ${\rm d}W=0$, and
there is one critical point for each basis tableau of $\cR_{n,k}$.
Since
\eq
     \det \left( {\partial^{2} W   \over \partial \hchi_{j} \partial \hchi_{k}}
              \right)  =
\det \left( {\partial^{2} W   \over \partial E_{j} \partial E_{k}}
              \right) =
{1 \over \Delta^{2}}
\det \left( {\partial^{2} W \over \partial q_{j} \partial q_{k}}
             \right) \; ,
\label{Hfirst}
\en
where (using eq.~\ref{symdef})
\eq
     \Delta =
          \prod_{i=1}^{n} q_{i}^{-1}
   ~\prod_{i=1}^{n} (q_{i} - q_{i}^{-1})
 ~\prod_{i < j} [(q_{i} + q_{i}^{-1}) - (q_{j} + q_{j}^{-1})] \; ,
\en
and since $\Delta$ does not vanish at the critical points,
the handle operator at the critical point corresponding to the
tableau $a$ is
\eq
   H(a) =  (-1)^{r(a)} [2i(n+k)]^{n}
              \left\{\Delta^{-2} \prod_{i=1}^{n}
      q_i^{-2} (q_i - q_i^{-1}) \right\}_{q_i=q_i(a)}\; .
\label{Hfin}
\en
This expression can be related to the $\sodnk$ modular transformation
matrices $S_{ab}$ as follows. The explicit form for the
modular transformation matrices\cite{kmod} for $B_n^{(1)}$ which
is usually given for $n\geq 3$ also holds for $n=1$
($A_1^{(1)}$) and $n=2$ ($C_2^{(1)}$). Therefore for all $n$ we have
\eq
     S_{ab}
  = (-)^{ n(n-1)/2 }
2^{n-1} (k+n)^{-n/2}  \det {\rm Mat}(a,b)
\en
with
\eq
   {\rm Mat}_{ij}(a,b)  = \sin
           \left( 2 \pi \theta_i (a) \theta_j (b) \over k+n \right),
\en
for $i,j = 1,\ldots,n$, and
\eq
  \theta_i (a) = \row_i (a) - i + n + \half, ~~~~~i=1, \ldots, n
\en
where the $\row_{i}(a)$ are the row lengths of the
tableau $a$ given in (\ref{dynkind}).

Then, since
\eq
    q_{j}(a) = e^{i\pi \theta_{j}(a)/(n+k)}
\label{auxcrit}
\en
are the values that the $q_j$ take at the critical point corresponding
to the tableau $a$,
we find that (with the $a$ dependence of the $q_{j}$ suppressed)
\eq
    S_{0a} = { (-1)^{n(n-1)/2} \over 2[2(n+k)]^{n/2} i^n }
   \prod_{i=1}^{n} (q_{i}^{\half} - q_{i}^{-\half})
 ~~\prod_{i < j} [(q_{i} + q_{i}^{-1}) - (q_{j} + q_{j}^{-1})] \; .
\label{Sone}
\en
In addition,
\eq
     \char_{\bsub{\bbox}}(a) =  {S_{\bbox a}\over S_{0a}}=
i^n (-1)^{r(a)} \prod_{i=1}^{n}
   {q_i^{\half} + q_i^{-\half}\over q_i^{\half} - q_i^{-\half}}  \; .
\label{bbchar}
\en
By combining (\ref{Hfin}) with (\ref{Sone}) and (\ref{bbchar})
the handle operator can be written
\eq
      H(a) = { 1 \over (\cS_{oa})^2}  {1\over \char_{\bsub{\bbox}}(a)}
\label{htwo}
\en
where $\cS_{0a} = 2 S_{0a}$ (\ref{SforR}).

Due to (\ref{detres}) the inverse operator of $\Phi_{\bbox}$,
\eq
    \Phi^{-1}_{\bbox} = \sum_b (\eta^{-1})_{0 b} \Phi_b \; ,
\label{inverse}
\en
exists as an integral linear combination of basis
elements of the chiral ring $\cR_{n,k}$.
The value of the field $\Phi_{\bbox}^{-1}$ at the $a^{\rm th}$ critical
point is given by one over the character at that point
\eq
      \Phi_{\bbox}^{-1}(a) = {1\over \char_{\bsub{\bbox}}(a)}
     = {\cS_{0a} \over \cS_{\bbox a}}    ~~.
\en
Therefore the handle operator satisfies
\eq
     H(a) =  (\cS_{0a})^{-2}   \Phi_{\bbox}^{-1}(a)
\label{hquest}
\en
for all critical points $a$.
Due to the non-singularity of the matrix $\cS_{ab}$
({\it c.f.}~\ref{Snorm})
this uniquely specifies the operators and we have
\eq
   H =  H_{0} \Phi_{\bbox}^{-1} ,
\label{uniform}
\en
where the untwisted handle operator $H_0$ satisfies
$$
 H_{0}(a) = (\cS_{oa})^{-2} \; .
$$
The tableau-basis expansion for the inverse operator begins
$\Phi_{\bbox}^{-1} = \Phi_{\bbox} + \ldots$ (to be compared with
$\Phi_{\bbox}^{-1} = \Phi_{\bbox}$ for
$\spnk$). The leading behaviour
is therefore $H=H_0 \Phi_{\bbox}$, as expected.

Due to (\ref{Snorm}) the correlation functions
are properly  normalized. For example, the residue
formula (\ref{residue}) and (\ref{hquest}) give
\eq
          \vev{\Phi_{a}}_{g=0} =
                  \sum_{\row_1(t) \leq k} \Phi_{\bbox}(t) \Phi_{a}(t)
                      (\cS_{0t})^2
= \sum_{\row_1(t) \leq k}  \cS_{\bbox t} \cS_{t a }
     =              \delta_{a \bbox }  \; ,
\en
where the sum is over all tableaux with $\row_1\leq k$.
Note that $\vev{\Phi_{\bbox}^{-1}} = \vev{\Phi_{\bbox}} = 1$
at genus zero. The topological metric for the theory based on the
$\sodnk$ superpotential is therefore given by
\eq
      \eta_{ab} = \vev{\Phi_{a} \Phi_{b}}_{g=0} = {M_{ab}}^{\bbox} \; ,
\en
as expected. This metric differs from the metric expected
on the basis of conformal perturbation theory in a flat basis,
$ \eta_{ab}^{0} = \delta_{b{\rho(a)}}$ (\ref{grassmet}).
In the $\sunk$ and
$\spnk$ cases the presence
of a simple current associated with the complement operation
led to the tableau basis being itself flat. Since
$\eta_{ab}$ is not anti-diagonal the three point function
\eq
       \vev{\Phi_a \Phi_b \Phi_c}_{g=0} =
     \sum_d {M_{ab}}^{d} \eta_{dc}
\en
does not collapse in this basis to a single fusion coefficient, as it did
in the $\sunk$ and $\spnk$ cases.

The correlation function of an arbitrary product of  $s$
chiral primary fields $\Phi_{a_{1}}$, \ldots, $\Phi_{a_{s}}$
on an genus $g$ Riemann surface
\eq
    \vev{\Phi_{a_1} \ldots \Phi_{a_s}}_{g} =
      \sum_{\row_1(t)\leq k}  {\cS_{0t}}^{-2(g-1)}
        { \cS_{a_1 t} \over \cS_{0t}} \ldots
        { \cS_{a_s t} \over \cS_{0t}}
\left(     { \cS_{\bbox t}\over \cS_{0t}} \right)^{-(g-1)}
\label{untwis}
\en
is almost certainly a non-negative integer.
Since the $\sodnk$ Verlinde numbers for a genus $g$ surface
\eq
     N^g_{\bsub{a \ldots z} } =
     \sum_{r}  {S_{0r}}^{-2(g-1)}
        { S_{a r} \over S_{0r}} \ldots
        { S_{z r} \over S_{0r}}
\en
(where the sum is over {\em all} integrable irreducible
representations of $\sodnk$, including spinors),
are known to be non-negative integers, we can prove this result directly
for $g=0$ and $g=1$. If $g=0$,
\eq
    \begin{array}{rcl}
        \vev{\Phi_{a_1} \ldots \Phi_{a_s}}_{0} & = &
\ds 4 \sum_{\row_1(t)\leq k}  {S_{0t}}^2
        { S_{a_1 t} \over S_{0t}} \ldots
        { S_{a_s t} \over S_{0t}}
        { S_{\bbox t}\over S_{0t}}   \\[0.6cm]
    &  = &
\ds   N^0_{\bsub{a_1\ldots a_s \bbox \hbox{\vphantom{$($}}   }} +
       N^0_{\bsub{a_1\ldots a_s \sigma(\bbox)}}
\end{array}
\en
and, of course, if $g=1$
\eq
       \vev{\Phi_{a_1}\ldots \Phi_{a_s}}_1 = N^1_{\bsub{a_1\ldots a_s}} \; .
\en
If $g>1$, then identity (\ref{inverse}) allows
an arbitrary $s$-point correlation
function on any genus to be written as a sum
of Verlinde numbers:
\eq
    \begin{array}{rcl}
  \vev{\Phi_{a_1} \ldots \Phi_{a_s}}_g & = &
\ds  {1\over 4^{g}}
\sum_{b_1} \eta^{-1}_{0b_1} \ldots  \sum_{b_{g-1}} \eta^{-1}_{0b_{g-1}}
\left( 4 \sum_{\row_1(t)\leq k} S_{0t}^{-2(g-1)}
         {S_{a_1 t}\over S_{0t}} \ldots  {S_{a_s t}\over S_{0t}}
 \prod_{j=1}^{g-1}{S_{b_j t}\over S_{0t}} \right)  \\[0.5cm]
 & = &
\ds  {1\over 4^{g}}
\sum_{b_1} \eta^{-1}_{0b_1} \ldots  \sum_{b_{g-1}} \eta^{-1}_{0b_{g-1}}
 \left( N^g_{a_1\ldots a_s b_1 \ldots b_{g-2} b_{g-1}}
 + N^g_{a_1\ldots a_s b_1 \ldots b_{g-2} \sigma(b_{g-1})} \right)
\end{array}~~~
\en
where the sum over each $b_i$ is over the
$n$ row tableaux with $\row_1\leq k$. While the Verlinde numbers
and the matrix elements $\eta_{0b_i}^{-1}$ are all
integers, it is not immediately clear
whether the correlation functions themselves must be integers due
to the fractional prefactor. We have verified the non-negative
integrality of many correlation functions
by explicit computation.
For example, since the one-point functions for $\cR_{2,1}$,
\setlength{\unitlength}{0.13cm}
\eq
   \begin{array}{ccccl}
 \vev{\Phi_{0}}^g & = &\vev{\Phi_{\fund}}^g & = & 3^{g-1}(2^g+(-1)^{g+1})\\
    &   &   \vev{\Phi_{\yfund{1.1}}}^g & = &  3^{g-1}(2^{g+1}+(-1)^{g})\; ,
\end{array}
\label{onepoint}
\en
are non-negative integers for all $g$, the correlation functions on any genus
are non-negative integers in this case.
The fact that the correlation functions are
(apparently) all non-negative integers suggests that these numbers
might have an interesting geometrical interpretation.

We expect the equality
$\vev{\Phi_{\bbox}^{-1}} = \vev{\Phi_{\bbox}}$, which follows
from (\ref{onepoint}) for $\cR_{2,1}$, to hold in general
(on arbitrary genus).

\subtopic{$\cN=2$ Rank-Level Duality}

Several works\cite{gepner,intri}
\setcounter{footnote}{1}
have commented on the manifest duality\footnote{Another kind of duality,
involving an interchange of the level of the $\cN=2$ algebra,
has been considered in ref.~\dcite{KS}.} of the Grassmannian
$\cG_{n,k}$ under interchange of $n$ and $k$. It was shown
in ref.~\dcite{brsym} that the chiral rings of the
$\spnk$ and $\spkn$ fusion superpotentials are isomorphic under
tableau transposition. (Rank-level-interchange identities
connecting the chiral rings of the $\sunk$ and $\sukn$ superpotentials
were also found in ref.~\dcite{brsym}, but they do not lead
directly to isomorphisms.)
Just recently, interesting implications of the rank-level
identities for $B_n$, $D_n$, and $C_n$ conformal weights
and modular matrix elements proved in ref.~\dcite{dual4}
have been found for $\cN=2$ superconformal theories.\cite{fs}

Here we show that the pair of topological field theories built by twisting
the $\cN=2$ models associated with the $\sodnk$ and $\sodkn$ \LG potentials
are also equivalent.  This follows from the
isomorphism $\cR_{n,k} = \cR_{k,n}$ given by tableau transpositon
(\ref{Riso}).  It also follows from the one-to-one correspondence
of primary fields of $\cR_{n,k}$ and $\cR_{k,n}$ given by transposition,
and the identity
\eq
    \left( S_{\vphantom{\wt{b\mathstrut}}a b}\right)_{\sodnk}=
  \left(   S_{\, \wt{a\mathstrut}\, \wt{b\mathstrut} } \,
       \right)_{\sodkn}  \; ,
\en
where $\widetilde{a}$ denotes the transpose tableau of $a$
(\ie, $\row_i(\ta)= \col_i(a)$).
This identity was proved in ref.~\dcite{dual4}.
The tableau language permits a uniform presentation of
rank-level-duality identities that relate each pair of
the $A_1^{(1)}$, $C_2^{(1)}$, and $B_n^{(1)}$ fusion rings
in various ways, as well as relating $C_2^{(1)}$ level two
with $C_2^{(1)}$ level two in a
different way than the $\spnk$ with $\spkn$ duality exploited
in ref.~\dcite{brsym}.
These results and expression (\ref{untwis}) imply that
\eq
   \vev{\Phi_{a_{1}} \Phi_{a_{2}} \ldots  \Phi_{a_{s}}}_{\cR_{n,k}}
 =  \vev{\Phi_{\ta_{1}} \Phi_{\ta_{2}} \ldots  \Phi_{\ta_{s}}}_{\cR_{k,n}}
\en
on Riemann surfaces of any genus.
Therefore the (twisted $\cN=2$) \LG models based on
the cominimally reduced $\sodnk$ and $\sodkn$ theories are exactly dual.

Even in the simple case of $\cR_{2,1}$ and $\cR_{1,2}$
the rank-level dual potentials look quite different:
\eq
        \begin{array}{lcl}
         \cV_{2,1}(x,y)& = &
{\sct {1 \over 4}} x^{4} + \half y^{2} - x^{2} y
+ x y - \half x^{2} + y - x   \\[0.2cm]
        \cV_{1,2}(x) & = &
  {\tx {1\over 4}} x^4 - x^3 + 2x  \; .\\[0.2cm]
         \end{array}
\en
Nevertheless, they lead to identical $\cN=2$ topological theories.

\topic{Comment on the Pattern}

The known (massive) $\cN=2$ \LG theories with chiral
rings related to WZW fusion rings fit a simple pattern:
They all arise as deformations that lead
from the quotient of a $U(N)$ representation ring
(the Grassmannian cohomology ring) that restores Young tableau
transposition symmetry
to  the quotient of a  $U(N)$--subgroup representation
ring that also maintains this symmetry. In addition, the tensor language of
the $U(N)$--subgroup Young tableaux gives a natural
description of the deformed chiral ring.
This pattern extends to all the classical
groups; no irregularity arises from the varying details of
the relevant Dynkin diagrams. It remains a (not impossible) challenge to
exhibit the exceptional groups as members of this pattern.
In the superconformal case, the quotient of the
$U(N)$  representation ring that restores the tableau
transposition symmetry has a beautiful, and manifestly
rank-level symmetric, geometric realization as the intersection ring of
Schubert cycles. It would be very interesting if an
analogous geometric interpretation could be found
for this symmetry in the massive case.

While we have not given direct evidence that the
$\sodnk$ related potentials yield integrable
theories of the kinks that interpolate between the
potentials' critical points, these theories should at least provide
an interesting test case for deciding which fusion ring properties
are necessary for integrability.
Although it might be tempting to suggest that
there is a one-to-one correspondence between the fusion
rings of rational conformal field theories and integrable
deformations,\cite{gepner,gs} the results presented here suggest that
there are likely more integrable deformations than just
those that reproduce conformal-field-theory fusion rings.

\vspace{0.3cm}
\noindent {\bf Acknowledgement}\hspace{0.2cm}
  We thank M. Bourdeau for many useful discussions.

\vspace{1.2cm}
\noindent {\bf Appendix}
\vspace{0.5cm}
\renewcommand{\theequation}{A.\arabic{equation}}
\onward

Parts one and two give the proof that the
transposition ideal $\cI_k$ contains the cominimal ideal $\cC_k$
and the entire fusion ideal $\cF_k$.  Part three contains the proof
that the extended Speiser algorithm implies those generators
of $\cI_k$ that are not instances of cominimal equivalence.
Part four compares this argument with the analogous
$\sunk$ and $\spnk$ arguments.

\vspace{0.3cm}
\noindent {\it 1. Proof that the transposition ideal contains
             all cominimal equivalence relations}
\vspace{0.4cm}

We will now show that the relations (\ref{genvan}) that
generate the transposition ideals $\cI_k$, which
read
\eq
    \tchi_{k+j} - \tchi_{k+1-j} = 0  \qbox{for} j = 1,\ldots,n \; ,
 \label{genvanrep}
\en
generate all of the cominimal equivalence relations (\ref{comin}).

Evaluation of the determinant formula (\ref{basdet}) for
the single-row character $\tchi_{k+j}$ yields the useful recursion
relation
\eq
    \tchi_{k+ j} = \sum_{i=1}^{n}[ (-1)^{i+1} \hchi_{i} (\tchi_{k+j-i}
          - \tchi_{k+j-2n-1+i})] + \tchi_{k+j-2n-1}~~~\qbox{for} k+j >
       2\; .
 \label{rec}
\en
If $j= n+1$ then the coefficient of each $\hchi_{i}$ vanishes
by assumption (\ref{genvanrep}) and relation
(\ref{genvanrep}) also holds for $j=n+1$.

An induction argument shows that (\ref{genvanrep}) also holds
for $n+2 \le j \le k+n$, as follows.
First, assume that (\ref{genvanrep}) holds
for all $j < m \leq k+n$.  Using this assumption
the recursion relation (\ref{rec}) for $j=m$ can be written
$$
    \tchi_{k+m} = \sum_{i=1}^{n}[ (-1)^{i+1} \hchi_{i} (\tchi_{k+1+i-m}
          - \tchi_{k+2-m+2n-i}) ] + \tchi_{k+2+2n-m}.
$$
Replacing $\tchi_{k+2+2n-m}$ by using (\ref{rec}) with $j=2+2n-m$,
and cancelling terms, yields $K_{k+m}-K_{k+1-m} =0$.
Therefore, by induction, we now know that
\eq
    \tchi_{k+j} - \tchi_{k+1-j} = 0  \qbox{for} j = 1,\ldots,n+k.
 \label{genplusk}
\en
While this is all that is needed in this section, we shall
need the case $j=n+k+1$ later.
If $n > 1$, the induction argument just given applies directly so
that equation~\ref{genplusk} also holds for $j=n+k+1$.  If $n=1$ (\ie, for
$\hbox{\sc so}(3)$), there is a slightly different result.
{}From (\ref{rec}) we find (for $n=1$, so that $j=k+2$)
$$
    \tchi_{2k+2} = \hchi_{1} (\tchi_{2k+1} - \tchi_{2k}) + \tchi_{2k-1}.
$$
Applying (\ref{genplusk}) to the right-hand side yields
$$
    \tchi_{2k+2} = \hchi_{1} (\tchi_{0} - \tchi_{1}) + \tchi_{2} = -K_0.
$$
To summarize, we have shown that
\eq
 \begin{array}{rll}
    \tchi_{k+j} - \tchi_{k+1-j} = 0&
 \qquad j = 1,\ldots,k+n, k+n+1  &\quad (n > 1)
       \\
    \tchi_{k+j} - \tchi_{k+1-j} = 0&  \qquad j = 1,\ldots,k+1 &\quad (n =1)
       \\
    \tchi_{k+j} + K_0 = 0& \qquad j = k+2   &\quad (n=1)
    \end{array}
 \label{onerow}
\en
(where, in fact, $K_0 =1 = K_{2k+1}$). Note that the equality in the
third line is not of
the form (\ref{genplusk}) and so is {\em not} a transpose of any of the
generators
(\ref{rmodgen}). (The same is true if $n>1$ with the first
deviation occuring for $j=k+2n$.)

The general cominimal equivalence relation (\ref{comin})
for any integrable representation $\tau$,
\eq
\char_{\bsub{\sigma(\tau)} } -
   \char_{\bsub{\hbox{\vphantom{$($}}\tau}} = 0           \; ,
\label{cominrep}
\en
now follows from the identities (\ref{onerow}) and the determinant formula:
Given an arbitrary integrable representation $\tau$
(so that $\row_1 + \row_2 \leq 2k+1$) the determinant in
expression (\ref{transdet})
for its character has first-row matrix elements
$$
\tchi_{\row_{1}+ j-1}+\tchi_{\row_{1}-(j-1)}  \qquad
              \rlap{$j=1,\ldots, \col_1$}
$$
which equal, due to (\ref{onerow}),
$$
      \tchi_{(2k+1-\row_1) + j-1} + \tchi_{(2k+1 -\row_1) -(j-1)}
    \qquad   \rlap{$j=1,\ldots, \col_1 \,  .$}
$$
The resulting determinant is exactly that of
the character of a representation with first-row length
$2k+1-\row_{1}$, but with all other row lengths the same as $\tau$.
This is exactly the definition of the action of $\sigma$ (\ref{tabcomin}),
so that the identities (\ref{onerow}) imply (\ref{cominrep}) for
an arbitrary integrable representation $\tau$.
Therefore, the transpositon ideal contains all cominimal equivalence
relations (for all $n\geq 1$).

\vspace{0.3cm}
\noindent {\it 2. Proof that the transposition ideal $\cI_k$ contains
             the fusion ideal $\cF_k$}
\vspace{0.4cm}

We begin by showing that (\ref{onerow}) implies a set of
{\em boundary} fusion relations for representations that just fail to be
integrable. Then we will show that these boundary
relations generate the entire fusion ideal $\cF_k$.

First consider an arbitrary representation with
$\row_{1} + \row_{2} = 2k + 2$.  For $n=1$, $\row_2=0$ and
the required relation is
just the last identity in (\ref{onerow}).
For $n\geq 2$, define $p$ for $k+2> p>0$ by setting
$\row_{1} = k + p$ and $\row_{2} = k+2-p$.
The determinant in (\ref{transdet}) for a representation
$\tau$ with first row length $\row_1= k+p$ has first row
matrix elements
$$
         \tchi_{k+p+j-1} + \tchi_{k+p-(j-1)}
\qquad   \rlap{$j=1,\ldots, \col_1\; .$}
$$
Since there are at most $n$ entries in this row ($\col_1 \leq n$),
the greatest subscript on $\tchi$, in any possible case, is $2k+n$.
Therefore, the identities (\ref{onerow}) imply that the matrix
elements of the first  row are
$$
\tchi_{(k+2-p) -2+j} + \tchi_{(k+2-p)-j}  \qquad
              \rlap{$j=1,\ldots, \col_1 \; .$}
$$
However, these are identical (in each column) to the
matrix elements of the second row of the determinant
so that the character must vanish:
\eq
      \char_{\bsub{\tau}} = 0
             \qquad {\rm for}\quad \row_1(\tau) + \row_2(\tau) = 2k+2
              \quad {\rm if} \quad n\geq 2\; .
\label{fusone}
\en

Now consider an arbitrary representation with
$\row_1 + \row_2 = 2k + 3$ (this case will only be needed for
$n\geq 2$).  Define $p$ for $k+2 > p>0$
by writing $\row_1 = k+1+p$ and $\row_2 = k+2-p$.
The matrix elements in the first two rows ($i=1$, $2$ below)
of the determinant (\ref{transdet}) for such a representation
$\tau$ are
$$
  \begin{array}{cc}
     \tchi_{k+p+j} + \tchi_{k+p+2-j}  & \quad i=1 \\
    \tchi_{k-p+j} + \tchi_{k-p+2-j}   &  \quad i=2
  \end{array}
\qquad   \rlap{$j=1,\ldots, \col_1 \; .$}
$$
Application of (\ref{onerow}) transforms these matrix elements into
$$
\begin{array}{cc}
    \tchi_{k-p+j-1} + \tchi_{k-p-j+1} & \quad i=1 \\
     \tchi_{k+p+j-1} + \tchi_{k+p-j+1} & \quad i=2
 \end{array}
\qquad   \rlap{$j=1,\ldots, \col_1 \; .$}
$$
Interchanging the first two rows, while leaving all others alone,
produces exactly the determinant for the character of a representation
$\overline{\tau}$ with
$\row_1 = k+p$, $\row_2 = k+1-p$, and all other row lengths equal to the
corresponding row lengths in the original tableau.  Therefore, the
original character ($\tau$) equals the negative of the
character ($\overline{\tau}$) obtained by removing
one cell from each of the first two rows of the original represention's
tableau, \ie
\eq
      \char_{\bsub{\hbox{\vphantom{$\overline{\tau}$}} \tau} }
               = - \char_{\,\bsub{\overline{\tau}}}
   \qquad {\rm for} \quad \row_1+\row_2 = 2k+3\; \quad {\rm and}
        \quad n\geq 2 \; .
\label{fustwo}
\en
A special case occurs when the second and third
row lengths in the original representation $\tau$ are equal.
Then $\overline{\tau}$ is not a valid tableaux and the
determinant produced above vanishes, so that $\char_{\bsub{\tau}}$ does too.

For $n=1$, the last identity in (\ref{onerow}) and,
for $n\geq 2$, the two identities (\ref{fusone}) and (\ref{fustwo}),
completely determine the fusion product of two arbitrary
basis elements $\phi_a$ and $\phi_b$ of the fusion ring, as follows.
First we expand $\phi_b$ as a polynomial in the $\hchi_i$.
Then associativity of the fusion product implies that
we can perform the product of
$\phi_a$ with a single $\hchi_i$ (for each term) first.
We compute this fusion product by first computing the tableau product.
For $n=1$ this corresponds to the tensor product between an
arbitrary tableaux and a single cell tableaux, which can only
increase the first row length by one, a case covered by
the last equation in (\ref{onerow}). (Note that, in terms of
$\su(2)$ tableaux, we are only dealing with even
length $\su(2)$ tableaux.)
For $n\geq 2$ this tableau product between an arbitrary
integrable tableau and a single-column tableau
can at most increase each of the first two row lengths by
one. Therefore, the only non-integrable tableau appearing are those with
$\row_1+\row_2 = 2k+2$, or $2k+3$, and these are exactly
the cases for which we can use (\ref{fusone}) and (\ref{fustwo})
to replace any such tableau with one with $\row_1+\row_2 \leq 2k+1$.
The iteration of this step
involving the fusion of a sum of (integrable) tableaux of the previous
step with single column tableaux results in the complete expansion of
the fusion product of $\phi_a$ and $\phi_b$ into a sum of $\phi_c$ with
$c$ an (integrable) tableau, without having to consider any non-integrable
representations with $\row_{1} + \row_{2} > 2k + 3$. Therefore, the fusion
ideal $\cF_k$ for $\sodnk$ is generated by the relations
satisfied by the two boundary cases
$\row_{1} + \row_{2} = 2k + 2$ and $\row_{1} + \row_{2} = 2k + 3$.

Now we will verify that the extended Speiser algorithm implies
the relations (\ref{fusone}) and (\ref{fustwo}) for $n\geq 2$ and
as well as the third line of (\ref{onerow}) for $n=1$.
Let $\varrho$ denote half the sum of positive roots for the algebra
being discussed.
If $n=1$, then the $A_1^{(1)}$ Dynkin index vector  for a highest weight plus
$\varrho$ is $[2K-2\row_1,2\row_1]$, since
$K=2k+1$ is half the usual $ \sun $ normalization, and since $a_1=2\row_1$.
If $\row_1=K+1$, a single Weyl reflection produces, after removal
of $\varrho$, $[0,2K]$, which is just the highest weight
of a single-row tableau of length $K$, as required by the third
line of (\ref{onerow}).

If $n=2$, then the $C_2^{(1)}$ Dynkin index vector for a highest weight
plus $\varrho$ is $[K+1-\row_1-\row_2, 2\row_2+1, \row_1-\row_2+1]$.
For $\row_1+\row_2=K+1$ this weight lies on the boundary of the
first Weyl chamber, and the character vanishes. For $\row_1+\row_2=K+2$,
then a single Weyl reflection produces
$[1, 2\row_2-1, \row_1-\row_2+1]$. If $\row_2=0$ then a further Weyl
reflection produces a weight on the boundary of the first Weyl
chamber and the character vanishes. If $\row_2>0$ then the vector
is positive and removal of $\varrho$ gives $[0,2\row_2-2,\row_1-\row_2]$,
which is just the highest weight of a tableaux with row lengths
$\row_i^{\prime}$ given by $\row_1^{\prime}=\row_1-1$
and $\row_2^{\prime} =\row_2-1$.
These are exactly the results expected for $n=2$.

If $n \geq 3$, the first case to consider is $\row_{1}+\row_{2}= 2k+2$.
The Dynkin indices of such a highest weight plus $\varrho$,
$$
(2k+2-\row_1-\row_2, \row_1-\row_2+1, \ldots, 2\row_{n}+1) \; ,
$$
give a weight which is on the boundary of the first Weyl chamber, so
that the character of any representation  with $\row_1 + \row_2 = 2k+2$
vanishes.

For the second case ($\row_{1}+\row_{2}= 2k+3$), let $\row_{1} = k+1+p$ and
$\row_{2} = k+2-p$, where $1 \leq p \leq k+1$.  The highest
weight Dynkin indices plus half the sum of positive roots are
$$
(-1, 2p, k+3-p-\row_{3}, \ldots)
$$
and we must perform the Weyl reflection
corresponding to the affine root to bring the weight into the
first Weyl chamber. The result
$$
(1, 2p-1, k+2-p-\row_{3},\ldots)
$$
has nonnegative entries.
The only possible zero occurs for
$\row_{3} = k+2-p$ (\ie, $\row_{3} = \row_{2}$), in which case we
again get a vanishing result.
Otherwise we obtain the Dynkin indices of an
integrable representation, $(0, 2p-2, k+1-p-\row_{3},\ldots)$,
corresponding to a representation with $\row_{1} = k+p$,
$\row_{2} = k+1-p$, and with all other row lengths the same as the original
representation.  Since an odd number of Weyl reflections were used,
the character of the new representation equals the negative of the character
of the original one. These results exactly reproduce (\ref{fustwo}),
including the special case in which the second and third row lengths
are equal.

The final conclusion is that the generating relations
of the ideals $\cI_k$ imply the fusion ideal generating relations
and all cominimal equivalence relations, as claimed in the text.

\vspace{0.3cm}
\noindent {\it 3. Proof that the extended Speiser algorithm and
         cominimal equivalence implies the generators of the
      the transposition ideal $\cI_k$}
\vspace{0.4cm}

For $n=1$ and $n=2$ the generators are instances of
the cominimal equivalence relations. For $n\geq 3$ the same is true
if $k+1\geq n$. For the remaining case $n\geq k+2$ we only need consider
the generators (\ref{genvan}) for $j=k+2, \ldots, n$, in which case
they read (since $k+1-j<0$)
\eq
         K_{k+j} =0        \qquad \rlap{$j=k+2,\ldots, n \; .$}
\label{finalcase}
\en
The highest weight vector plus half the sum of positive roots
for $K_{k+j}$ is
$$
   [k+2-j, k+j+1, 1,\ldots, 1] \; .
$$
Since the $(j-k-1)^{\rm th}$ entry vanishes after $j-k-2$ Weyl
reflections for each $j=k+2,\ldots, n$, all of these weights are on the
boundary of a Weyl chamber, as required by (\ref{finalcase}).

\vspace{0.3cm}
\noindent {\it 4. Comparison with the $\sunk$ and $\spnk$ cases.}
\vspace{0.4cm}

It is instructive to compare the above arguments
with the analogous arguments for $\sunk$ and $\spnk$.
The tableaux of an integrable $\sunk$ or $\spnk$ representation
satisfy $\row_1 \leq K$.  The appropriate determinant formula
is just that given in (\ref{grassdual}) for $\sunk$
(with the $\hchi^{\sigma}_i$
taken to be the characters of single-column tableaux and a different
set of rank modification rules imposed) and
just that given in (\ref{basdet}) for $\spnk$.
Using these determinant expansions (and the appropriate
rank modification rules) for  arbitrary basis fields as
polynomials in single-column tableaux,  the fusion product of
two representations only requires computing a series of fusion products
of general integrable representations with single-column tableaux.
Hence, any {\em non}-integrable tableau occuring at any step in this process
will have $\row_1 = K+1$.  We know from the extended Speiser
algorithm that all such tableaux have vanishing characters.
In order to find a set of generators which imply that every representation
with $\row_1 = K+1$ has vanishing character, consider
the row-character expansions (\ref{fgrassdet}) for $\sunk$
and (\ref{transdet}) for $\spnk$. Since the rows beyond the first of
any such vanishing determinant
must allow arbitrary variations (according to the arbitrary
values of the lower rows of the initial tableau), the only way every
such determinant can vanish is for the top row of the determinant
to also vanish. Therefore, these top-row-entry
vanishing conditions (in terms of the characters $\colK_j$ of the
single-row tableau with $j$ boxes)
$$
\colK_{K+1+j} + \delta_{G} \colK_{K+1-j} = 0
          \rlap{$\qquad$ {\rm for} $j= 0,\ldots, n-1\; ,$}
$$
where $\delta_{G}$ is zero for $\sunk$ and one for $\spnk$ and
$n$ denotes the rank of the relevant group,
generate the conformal-scalar fusion rules of the relevant current algebra.

\end{document}